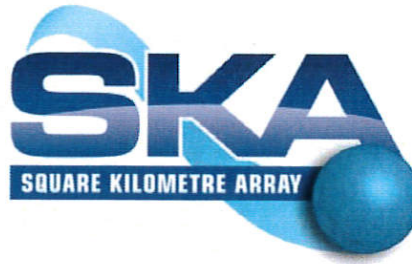

**SKA CSP MEMO 0021**

Note: this document is only for historical reference and is not a reference for the current SKA Mid telescope design.

## Sample clock frequency offset (SCFO) Resolution Team 3 investigation

# *SKA Mid Telescope, RT-3 REPORT*

**Author(s)**: B. Carlson (CSP: NRC-Canada), P. Boven (SADT: JIVE-Netherlands), K. Caputa (DISH: NRC-Canada)

**Date**: 2016-04-18

**Revision:** 1

| **Prepared By** | Name | Brent Carlson | Signature Date | Brent Carlson (Apr 20, 2016) |
|---|---|---|---|---|
| | Organisation | NRC-Canada | | |
| **Reviewed By** | Name | Paul Boven | Signature Date | Paul Boven (Apr 26, 2016) |
| | Organisation | JIVE | | |
| | Name | Kris Caputa | Signature Date | K. Caputa (Apr 27, 2016) |
| | Organisation | NRC-Canada | | |
| **Approved By** | Name | Gie Han Tan | Signature Date | G H Tan (Apr 28, 2016) |
| | Organisation | SKAO | | |
| **Issued By** | Name | Brent Carlson | Signature Date | Brent Carlson (Apr 20, 2016) |
| | Organisation | NRC-Canada | | |

# ABSTRACT

*This Resolution Team 3 (RT-3) report contains the results of the investigation of key aspects of the proposed **Sample Clock Frequency Offset** (**SCFO**) scheme for the Mid SKA1 telescope. This is a scheme, first proposed by one author (Carlson) at a meeting at the SKAO 25-Jan-2013, to digitize the analog signal at each antenna at a slightly different sample rate and transmit the data to the CSP Mid.CBF for subsequent digital re-sampling to a common sample clock frequency before channelization, correlation, and beamforming. The primary purpose for doing this is to cause de-correlation of sample clock-related self-interference to be able to improve correlated and beamformed signal quality. This report includes an investigation of the efficacy of the method, investigation of its implementation by SADT, DISH, and CSP, presentation of further supporting modeling results augmenting the original modeling work present in* [1]*, a note on expansion to SKA-2, as well as possible draw-backs and concerns. Finally, there are potential additional benefits in signal quality in implementing the SCFO scheme, in particular de-correlation of aliased RFI (particularly for Nyquist Zone-2 digitized signals) as well as relaxation of signal-chain anti-aliasing filters' transition band roll-off and reject band attenuation prior to digitization in the antenna.*

# Table of Contents

# 1   Introduction and Purpose

The purpose of this report is to provide background investigation material to support an ECP to implement the Sample Clock Frequency Offset (SCFO) scheme for the Mid SKA1 telescope.  Three SKA1 Elements are primarily impacted, must be engaged to implement this scheme, and therefore contribute to this report.  The impacts/issues for each Element are summarized below:

- SADT – Must deliver to DISH the required coherent sample clock frequencies, each different (with at least ~100 Hz resolution) by potentially up to ~+/-1 MHz[1] for each dish (+/-10 MHz desired), as well as allowing for an offset frequency of 0 Hz (i.e. no offset).  Along with these frequencies is the 1PPS (1 pulse per second) time marker used by the DISH digitizer to timestamp sampled data transmitted to the CSP.  It is agreed that the sample clock used at the digitizer does not need to be an integer Hz as long as the frequency *offset* is precisely known (and expressed as a rational number); phase is not known but is established at the start of an observation using a sky calibrator and the clock is free-running afterwards.  This is because the offset frequency is generated by a DDS (Direct Digital Synthesizer) that has frequency resolution limitations; additional phase steering on top of the DDS could be employed but it would add unwanted and unacceptable phase noise.

- DISH – Must be able to use "slightly" different sample clock frequencies at each antenna and a) that there are not an integer number of clock cycles between 1PPS ticks and b) that the 1PPS is not synchronous with the sample clock.  DISH should therefore not depend or rely on any particular number of clock cycles between 1PPS marks.  This condition can be relaxed if necessary, e.g. multiples of 100 clock cycles (i.e. 100 Hz tuning resolution).  DISH generates a digitized data stream to CSP with 1PPS markers (timestamp), but CSP cannot rely on there being the same number of samples between any pair of 1PPS and subsequent markers because a) the sample clock frequency is not an integer Hz frequency and b) the 1PPS— delivered to DISH by White Rabbit—is not synchronous with the SADT-delivered and phase-stable clock at DISH.

- CSP – Must re-sample the digitized data to a common sample rate prior to correlation and beamforming.  Such re-sampling must therefore be realizable and implemented without an unreasonable impact on signal fidelity, cost, or power of the CSP.  The SCFO scheme has no known impact on beamformed data products and subsequent PSS and PST data product fidelity, based on first principles, as will be shown later.

Other SKA1 Elements are also impacted, but to a far lesser degree and are therefore not directly involved in the preparation of this report.  These are noted as follows:

- TM – Must provide SADT, DISH, and CSP with each dish's absolute sample clock frequency, or offset frequency from the established sample frequency for each Band.  An alternative to this is that there could be a built-in algorithm in SADT, DISH, and CSP where the offset frequency is the product of the dish number and a base offset frequency.  For maximum flexibility (i.e. taking into account sub-arraying, the desired spectral artefact suppression effect, integration time, dish number, Band, and observation science goals) it is best that TM explicitly provide the sample clock offset frequencies from the per-Band established common frequency.

- SDP – There should be no impact on the SDP and it should not need any knowledge of the implementation.  This is assuming that the offset frequencies are not impinging on the spectrally-pure visibilities received from CSP, a function of the offset frequencies and the

---

[1] Allowing 10 kHz minimum baseline differential frequency offsets on 200 antennas, resulting in ~40 dB suppression in a 0.14 sec integration time.

> original oversampling factors (i.e. knowing that a significant fraction of the sampled band— guard band—is throw-away anyway).

Since the SCFO scheme relies on there being a significant fringe washing effect for every correlated baseline, and the amount of washing is proportional to the product of the frequency difference (i.e. between the frequencies used at each antenna) and the correlator integration time, it is generally advantageous for the frequency difference be as large as possible. For a large array, the worst-case baseline frequency difference can be large. What this means is that the SCFO scheme is only practical if applied to wideband digitization and removed prior to any subsequent narrower-band channelization.

Finally, it is worth noting that the EVLA and e-MERLIN telescopes and "WIDAR" correlators successfully used a similar method—mixer LO frequency offsets—for the purposes of anti-aliasing [2] (with the benefit also of washing out digitizer self-interference [3]). In those correlators a 3-level baseline-based fringe rotator was used and it was found that proper selection of offset frequencies (i.e. prime numbers) was important to ensure that the 3-level phase function was properly "repetitively" sampled across many cycles of phase within a correlator integration time. Not doing so resulted in low-level amplitude modulation of the visibilities. For the CSP Mid.CBF system no such coarse phase rotator will be used—all phase rotators will use many bits of precision (at least 12, determined by the size of the sine/cosine memory LUT) and so this is not an issue. Even so, without the SCFO scheme, such mixers are required anyway to implement phase delay and earth-rotation phase correction.

# 2 Overview

This section provides an overview description of the SCFO scheme as well as a total SADT-DISH-CSP critical signal flow diagram and description.

## 2.1 SCFO Scheme

The SCFO scheme is summarized in the following bullets:

1. Digitize the wideband analog signal with a different sample rate at each antenna, where the worst case (largest) differences are a small fraction of the digitized bandwidth[2]. Make the sample rate at each antenna as different as possible. Ensure that no frequency that is common to every antenna leaks into any part of the analog signal path.

2. *Any* signals that leak into the analog signal path with antenna sample-rate dependency will not correlate since they are imprinted into the analog signal at different frequencies at different antennas.

    The amount of correlation ρ between two different frequencies, $\omega_1$ and $\omega_2$ fundamentally is:

    $$\rho = \frac{1}{T} \cdot \int_0^T \sin(\omega_1 t) \cdot \sin(\omega_2 t) \, dt$$

    Which is the fringe washing function, sinc(($\omega_1$-$\omega_2$)T), with envelope 1/(($\omega_1$-$\omega_2$)T). For example, for T=1 second and a 1 kHz frequency difference, the suppression effect is ~38 dB.

3. Prior to channelization and correlation, the digitized signal is "re-sampled" to a common sample rate (i.e. one that is the same for each dish) using an interpolation filter. This re-sampling operation is equivalent to converting the digitized signal to an analog signal and then re-digitizing it with the common sample clock *except* there is no analog conversion and so there is no possibility for signal contamination with the common sample clock.

4. If the original signal is digitized at other than Nyquist Zone 1 (i.e. baseband), then a sample frequency-dependent down conversion has occurred (i.e. the down conversion is implemented with the digitizer). Since the sample clock frequency is different at each antenna a different down conversion has occurred and the resulting signals are at different frequencies. These different frequencies must be removed prior to channelization and correlation with a frequency shift/mixing operation.

5. For Nyquist Zone 2 sampling, aliased signals in all reject bands outside the pass-band won't correlate since their frequency shift is different than the pass-band frequency shift. For Nyquist Zone 1 sampling, there is a similar suppression effect, but some regions outside the pass-band will correlate—see section 3 for details.

A "proof by pictures" of the sampling, re-sampling, and correlation process outlined above is presented in section 3.

For beamforming there is no de-correlation/washing effect since correlation is not performed. However, what happens when the re-sampled signals from multiple antennas are added together in the beamformer is that they do not add coherently. This improves the SNR of the beam output since the desired astronomical signal adds coherently, whereas spectral artefacts and noise do not add coherently.

---

[2] And normally don't "cut into" the correlated bandwidth delivered to the SDP. i.e. there will be de-correlation at the band edges with the de-correlated bandwidth equal to the frequency difference. See Figure 3-2.

## 2.2 SADT-DISH-CSP Signal Flow Diagram

A simplified overview signal and timing flow diagram of the SADT-DISH-CSP timing path is shown in Figure 2-1. Some items of note in this figure are as follows:

1. The red lines (and boxes) are coherent clock regioins where absolute phase stability is of critical importance to maintain coherence of the system. The blue lines are all discrete-time or discrete-time-oriented and so such absolute phase stability is much less important, although all clocks must be locked to the maser to ensure continuous synchronized processing.

2. One of 2 methods may be employed for offset sample clock distribution to the antennas. In Figure 2-1 (a) SADT delivers "fc"—the common sample clock frequency—to a synthesizer in the antenna pedestal via a round-trip, phase corrected method. In Figure 2-1 (b), SADT delivers "fa"—the antenna sample clock including the offset frequency to each antenna. In (a) an "offset frequency synthesizer" in the antenna generates "fa", the antenna-specific sample clock frequency. In this case it is critically important that fc does not leak into the antenna's analog signal in any way since such leakage will correlate. In (b) the fa frequency can be used by the antenna directly for its sample clock (or multiplied up accordingly if needed).

    In both these cases, the 1PPS pulse delivered to the antenna by SADT using "White Rabbit" is constrained to be within a few nsec of the ideal, but its precise phase relative to fc (a) or fa (b) is unknown. The DISH digitizer will sample this pulse and deliver it to CSP such that is has the same degree of accuracy but that the number of samples between consecutive 1PPS ticks will not be the same (i.e. vary around some average). Indeed, since SADT (or DISH) uses a DDS for SCFO generation, the 1PPS tick at DISH will be drifting relative to fa (whether generated centrally or locally) since fa has a rational frequency but not in integer Hz. Nevertheless, each 1PPS pulse in the antenna will be sampled and sent to the CSP to establish and maintain required timing accuracy in the system.

3. The 100 MHz discrete-time quality clock delivered to the CSP by SADT is used by CSP to synthesize fc—the common sample clock—for its internal use. As shown in section 6, at no time does CSP actually have to produce any antenna sample clock fa, although it does have to synthesize fc-fa for each antenna processed. The 100 MHz and 1PPS at the SADT-to-CSP interface are locked together and synchronous such that there are always $100 \times 10^6$ clock cycles between 1PPS ticks. This allows Mid.CBF to establish a rigorous and predictable time standard which—during a 0 delay initialization phase at the beginning of each observe block[3]—will be referenced to the 1PPS ticks coming from the antennas to establish absolute timing accuracy (for timestamping of data and application of delay/phase tracking models) to the accuracy of the 1PPS pulse relative to the ideal 1PPS at each antenna.

Further detail of SADT, DISH, and CSP implementation is contained in following sections.

---

[3] Defined for use in this report as a sequence of calibration and science observe "scans" where SDP must connect phases on all baseline visibilities from calibration-to-calibration scan across science scans leading to one coherent image.

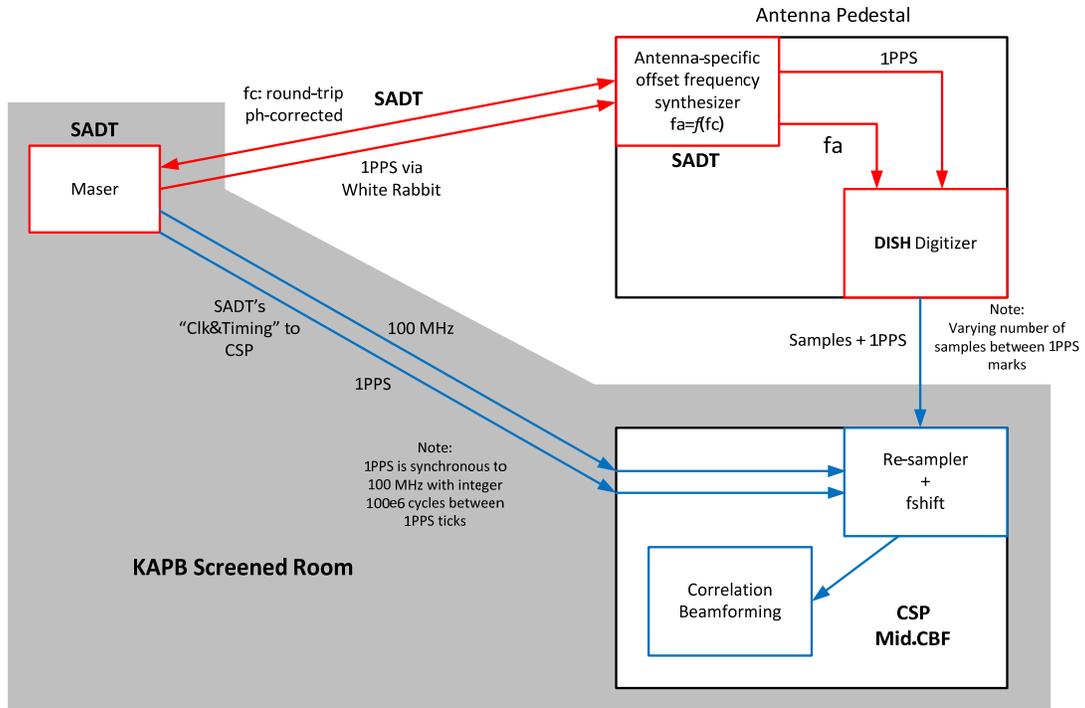

(a)

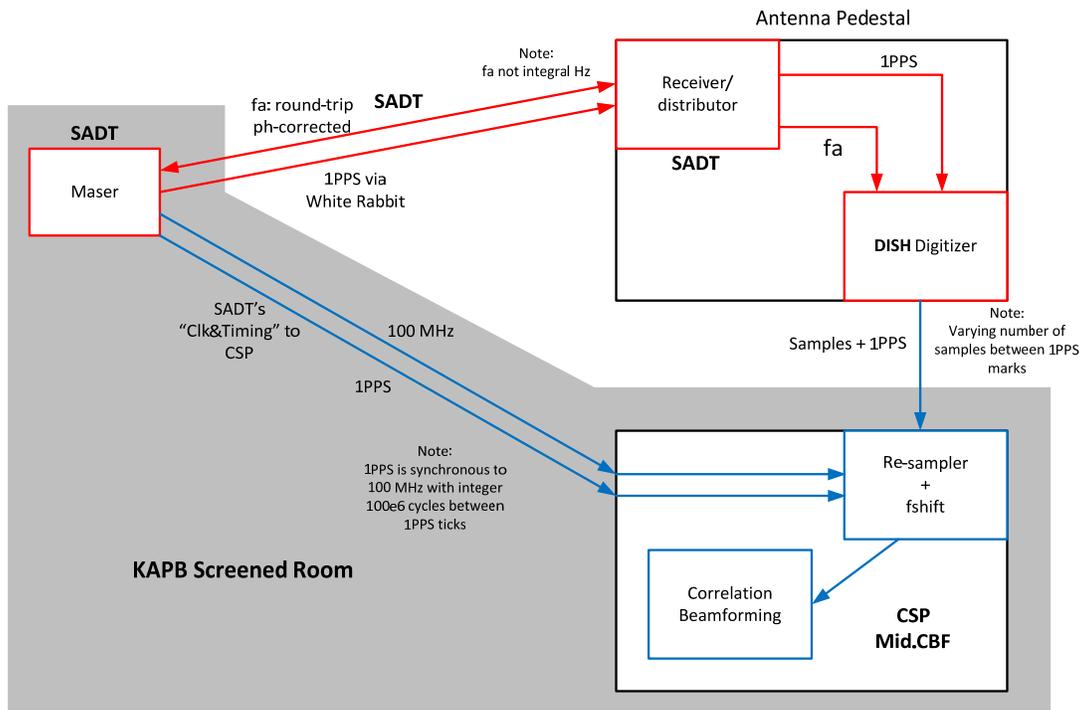

(b)

*Figure 2-1  Simplified signal and timing diagram of the SADT-DISH-CSP timing path.  In this diagram, red lines indicate high phase stability coherent clocks; blue lines indicate discrete time clocks or data with much less stringent phase stability, but of course, sourcing from the Maser for continuous synchronous processing.  In (a) SADT sends the common clock fc to each antenna; in (b) SADT sends a different clock to each antenna.*

# 3  SCFO Scheme Explained: "Proof by Pictures"

This section contains a more detailed explanation of the SCFO scheme and presents a "proof by pictures" of the scheme using frequency convolution diagrams. It also uses the same method to illustrate the additional benefit of aliased out-of-band spectral contamination suppression.

Figure 3-1 shows the initial digitization and re-sampling frequency convolution diagram for "antenna 1". The red interference lines are self-interference and are sample clock 1-derived frequencies. The blue interference lines are from the other "antenna 2" leaking into the antenna 1 analog signal. In this case the antenna 1 sample clock is < the common sample clock and the resulting (exaggerated) aliasing of the signal is shown—in practise, fa can be > or < fc and this aliasing will be within the normally unusable transition band anyway so is of no consequence.

A similar diagram was developed for antenna 2, but is not shown.

In Figure 3-2, the steps of frequency shifting and correlation of the re-sampled antenna 1 and antenna 2 signals are shown. Note that the self-interference introduced into each antennas' signal (antenna 1—red lines, antenna 2—blue lines) doesn't correlate, but the leaking cross-interference does. This, of course, is expected—any common signal showing up in the pass-band in both antennas correlates.

What these results do mean, though, is that sample-clock dependent self-interference doesn't correlate and so employing the SCFO scheme can mean:

- Filtering, shielding, and handling of the antenna-dependent sample clock can be relaxed somewhat but not so much that it leaks into an adjacent antenna's receiver. This includes leakage of the sample clock (including any of its harmonics/sub-harmonics) into the analog path within the A/D converter module/device itself, often a very difficult problem to deal with.

- Antenna-dependent sample clocks can be distributed within the antenna at lower frequencies knowing that those frequencies won't correlate if they leak into the signal path. This can ease sample clock distribution where the final/highest sample frequency is only produced where it is needed.

- Interleaved samplers can more comfortably be used since the sample rate-dependent spectral "birdies" introduced/imprinted into the signal due to interleaving won't correlate. This can open up a whole new acceptable sampling technology since often the highest-performance/highest-bandwidth samplers are typically interleaved devices and interleaved devices, no matter the device-based mitigation approach, produce artefacts that will correlate.

The signal flow examples of Figure 3-1 and Figure 3-2 use Nyquist Zone 2 sampling. If Nyquist Zone 1 (baseband) sampling is used, the final frequency shift is not needed but the sample clock self-interference still does not correlate because, again, the self-interference is a function of the antenna-base sample clock and therefore is at a different frequency for each antenna.

**Out of Band Spectral Contamination Suppression**

As briefly mentioned in section 2.1, for Nyquist Zone 2 sampling, spectral contamination at frequencies in the spectrum outside of the pass-band of interest do not correlate. This can mean that out-of-band interference and "ghost" spectral images, however and wherever produced and that alias into the correlated band due to digitization, won't correlate. This can mean that analog band-pass filter reject-band requirements can be relaxed and that analog gain compression producing spectral images that alias into the pass-band won't correlate.

Figure 3-3 to Figure 3-5 illustrate for Nyquist Zone 2 sampling how out-of-pass-band spectral artefacts—whether related to the antenna sample clock frequency or not, and wherever located, do not correlate.

Figure 3-6 is the same out-of-pass-band spectral contamination as Figure 3-3, except for Nyquist Zone 1 (baseband) sampling.  Note that in this case one spectral feature does correlate and, indeed, all spectral features in the indicated region outside of the pass-band, will correlate.  This and the preceding diagrams illustrate that Nyquist Zone-2 sampling is advantageous to maximize out-of-band unwanted spectral feature suppression.

Refer to [1] and section 7 for modelling of re-sampling including sensitivity losses due to a) the number of interpolation filter taps, b) the number of interpolation steps between +/-0.5 samples and, c) additional sensitivity losses from re-sampling initially digitized signals that are coarsely quantized in amplitude.  Regarding the latter, [1] indicates there is some uncertainty in sensitivity losses when interpolating a coarsely-quantized signal.  Section 7 clears this up for the worst case of 4-bit samples in Band 5—i.e. additional sensitivity loss for 4-bit sampled data is of little consequence.

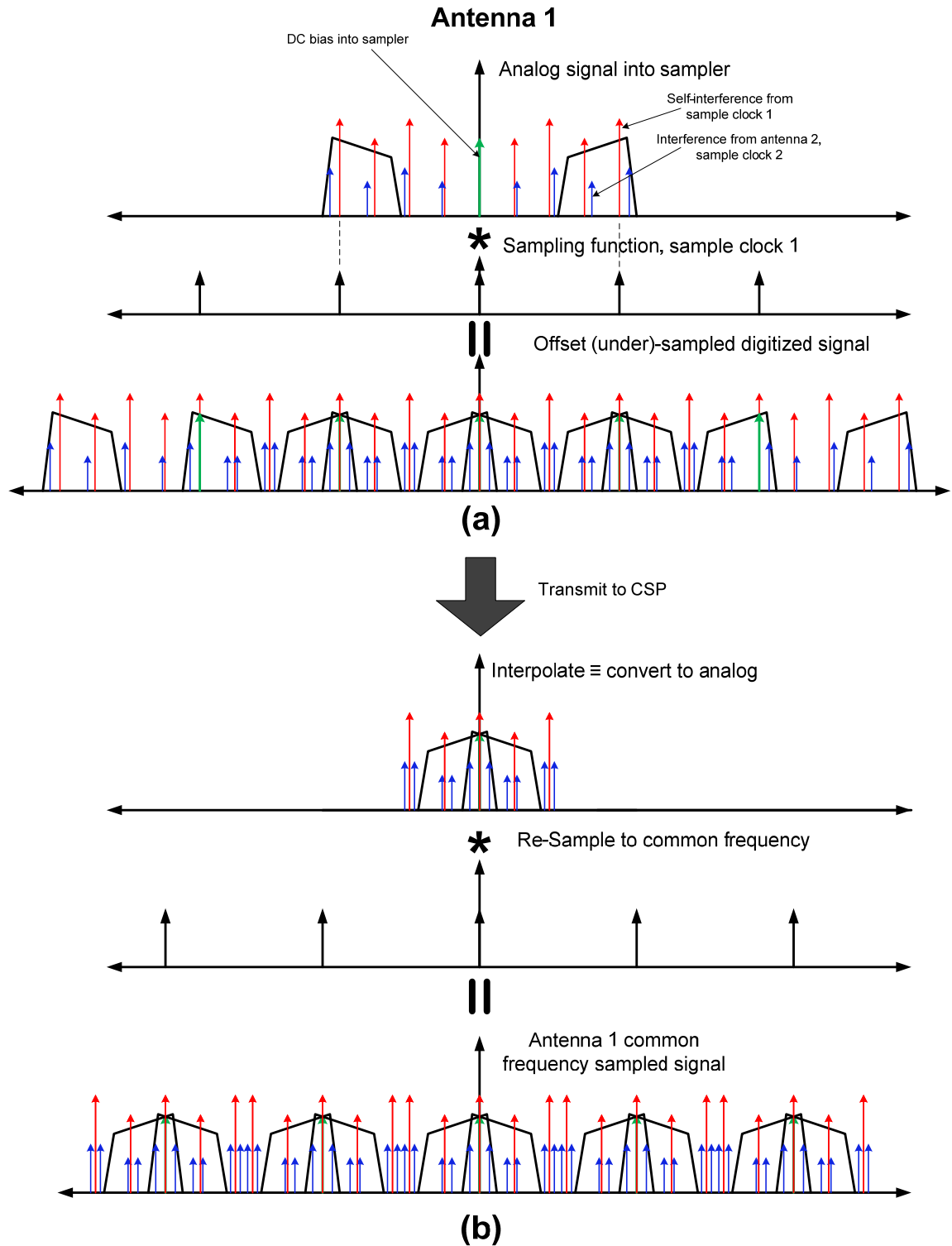

*Figure 3-1 Antenna 1 initial digitization (a) and re-sampling (b) frequency convolution diagrams. The red lines are self-interference at sample clock 1 derived frequencies. The blue lines are cross-interference from the other "antenna 2" at sample clock 2 derived frequencies.*

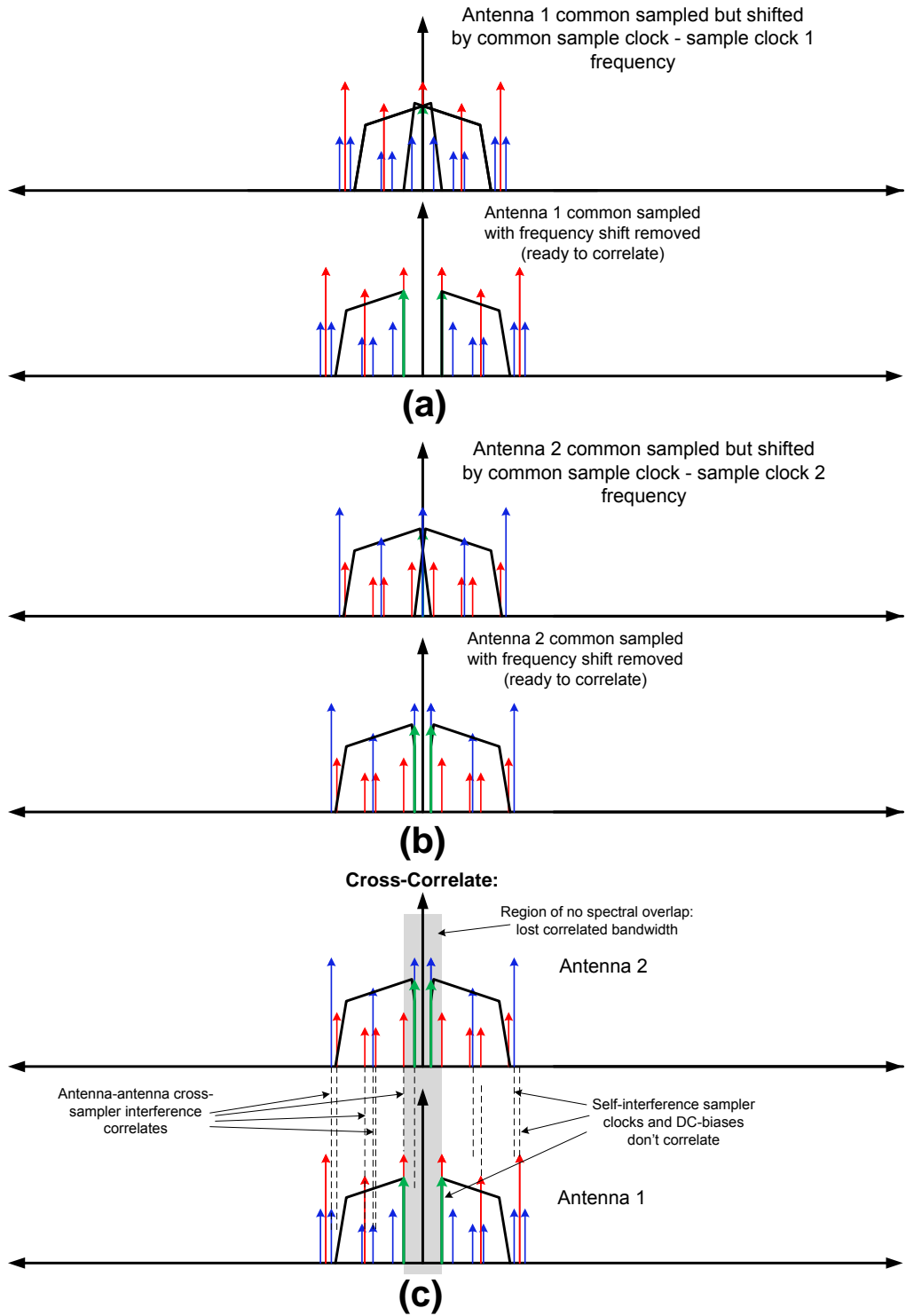

*Figure 3-2  Final frequency shift removal, (a) antenna 1, (b) antenna 2, and (c) correlation.  The sample clock self-interference signals (blue arrows in antenna 2 and red arrows in antenna 1) don't correlate.  For clarity copies of the signal across the spectrum, present in a digitized signal, are not shown.*

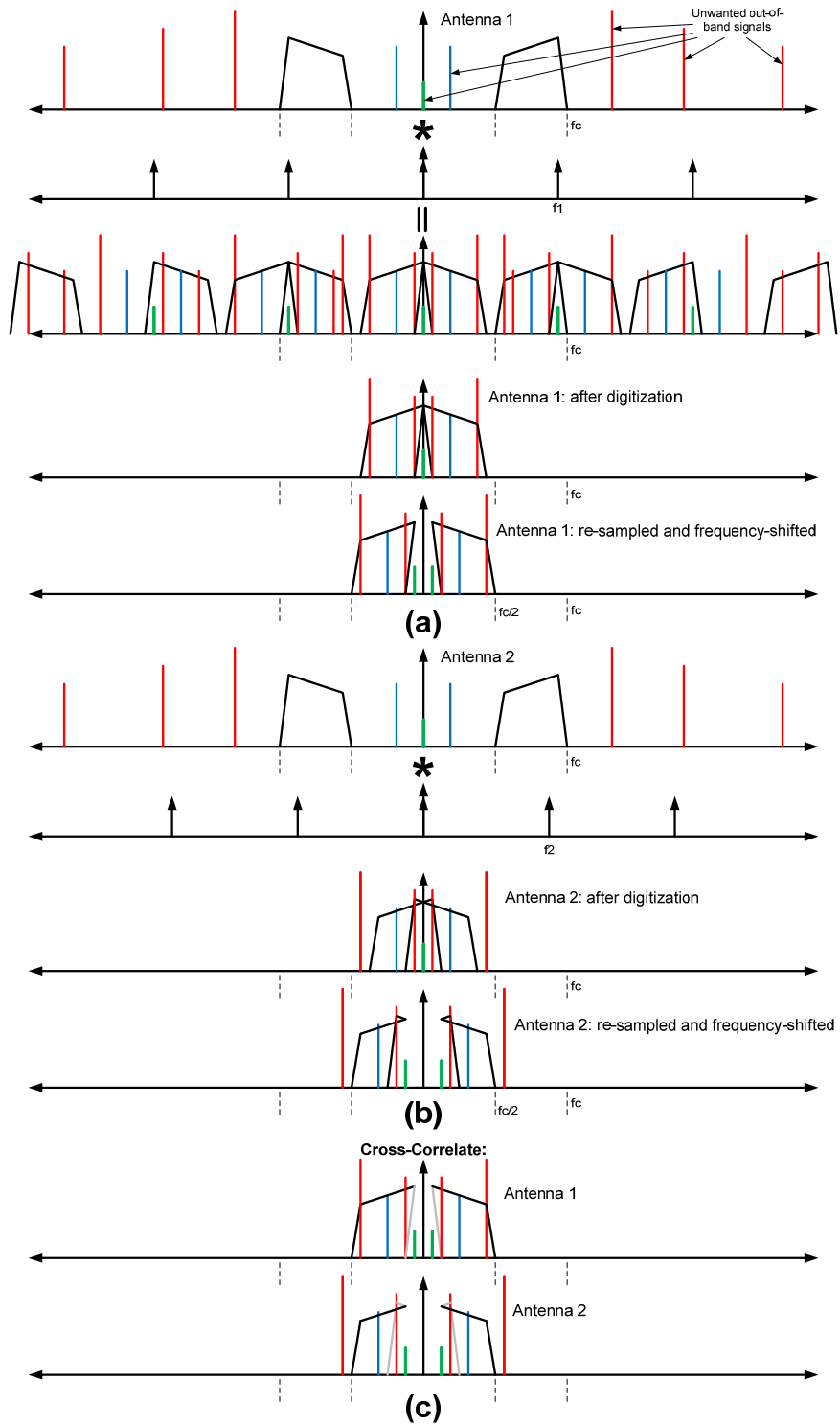

*Figure 3-3  Convolution diagrams illustrating that out-of-pass-band signals, not originally related to any sample clock, do not correlate—Nyquist Zone 2 case.  For clarity, in most cases post-digitizer spectral copies are not shown.  The black lines in (c) indicate the pass-band, which does correlate.*

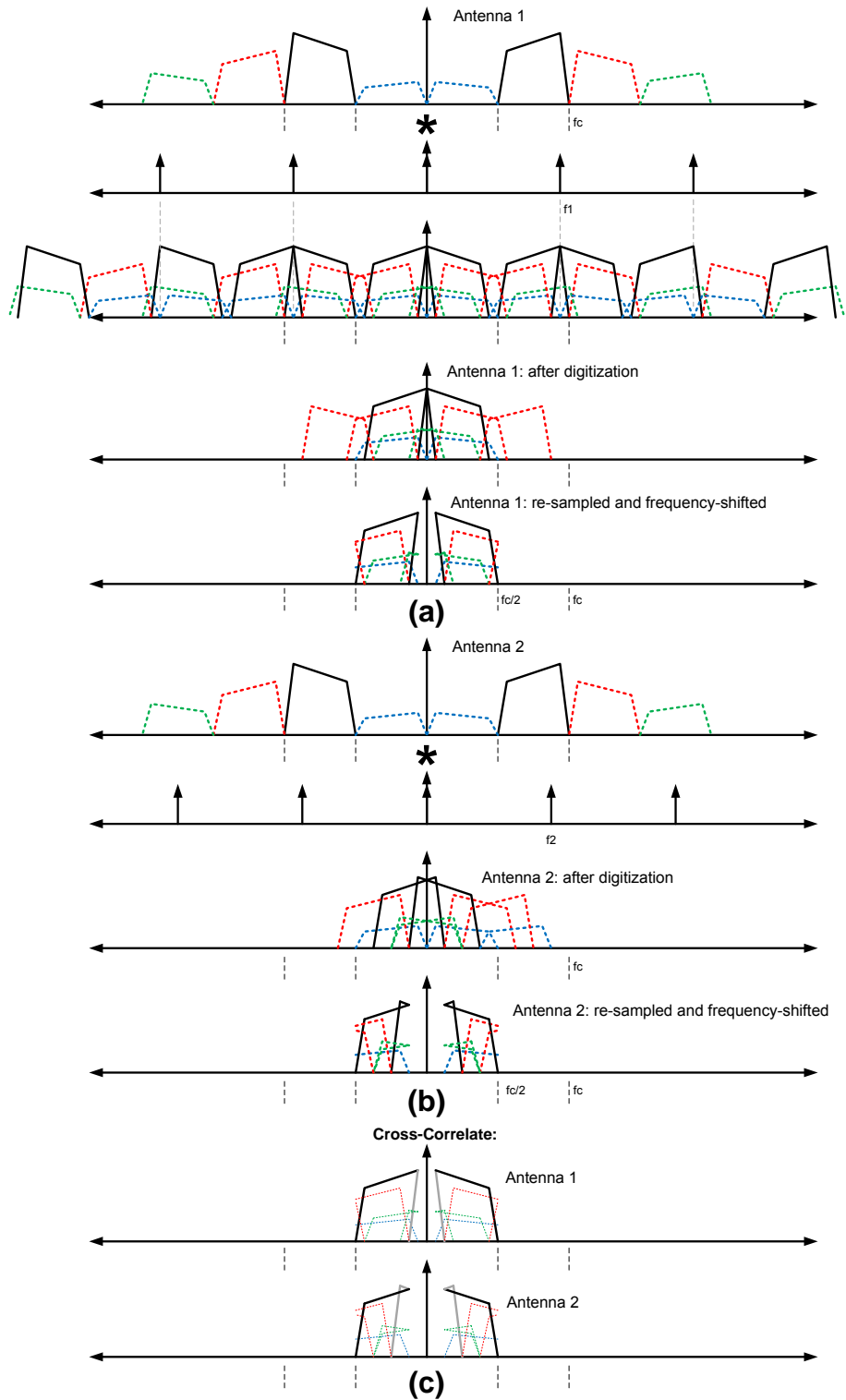

*Figure 3-4  Convolution diagrams illustrating that out-of-pass-band signals, related to antenna sample clock frequencies, do not correlate—Nyquist Zone 2 case.  For clarity, in most cases post-digitizer spectral copies are not shown. The black lines in (c) indicate the pass-band, which does correlate.*

The last portion of Figure 3-4 is difficult to see.  It is reproduced in 2 views, (a) and (b) in Figure 3-5:

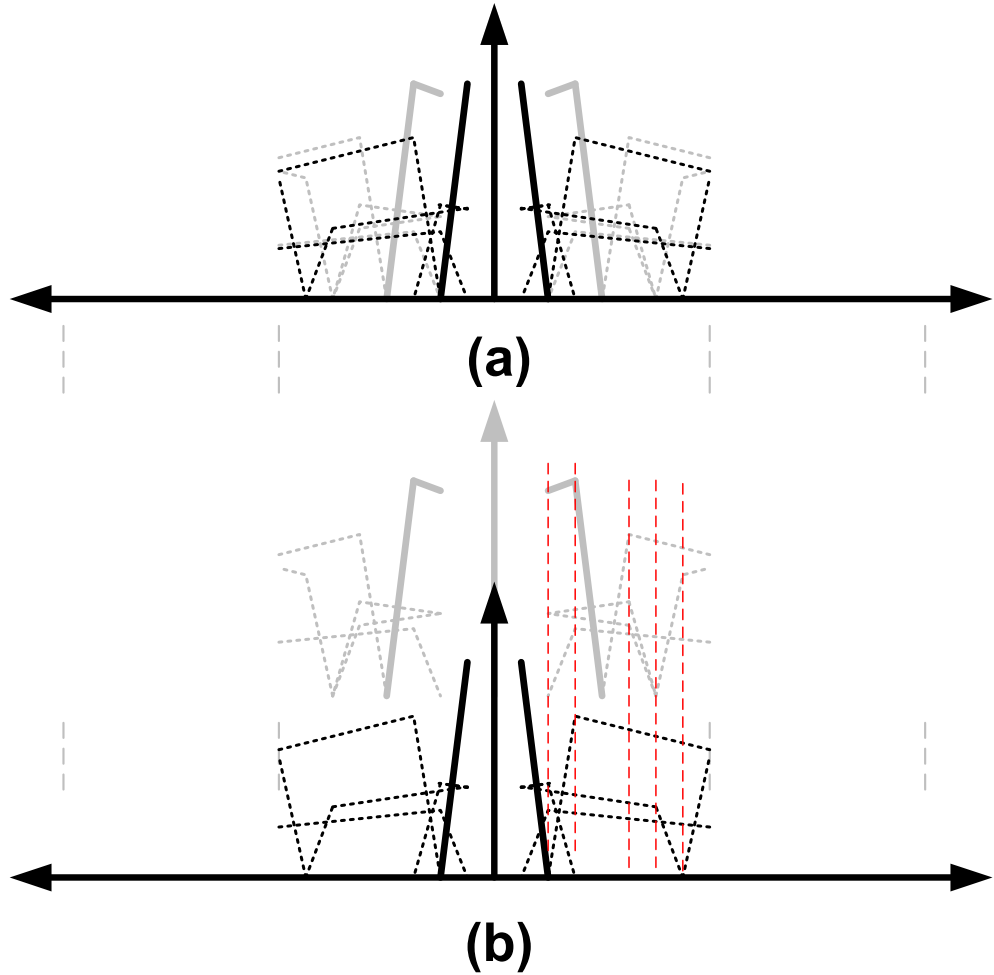

*Figure 3-5  Aliased artefacts from Figure 3-4 not showing the pass-band (the part of the spectrum that does correlate), ready for correlation—Nyquist Zone 2 case.  Antenna 1 (black) artefacts and antenna 2 (grey) artefacts (both in (a) and (b)) do not line up in frequency and therefore do not correlate.*

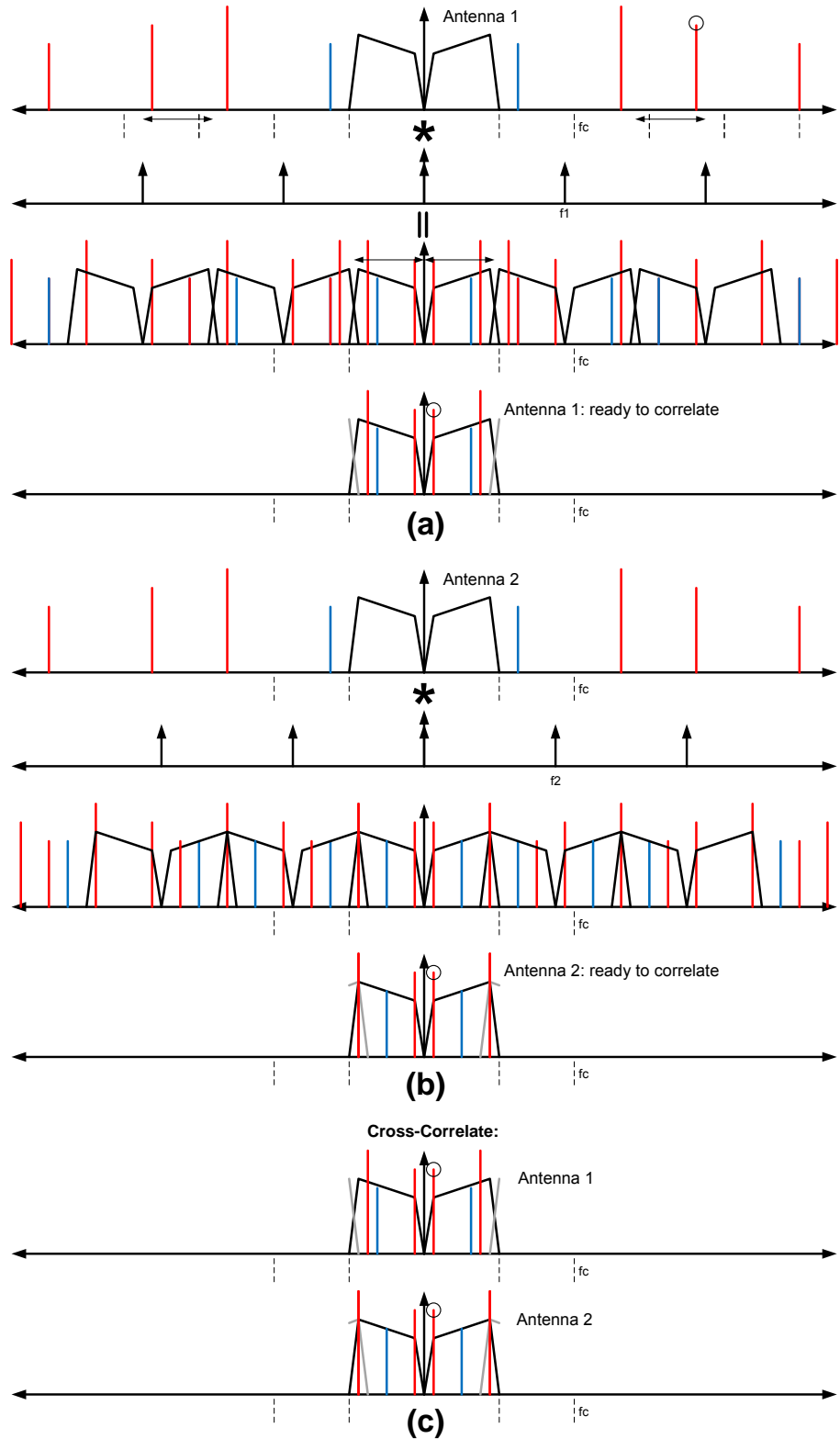

*Figure 3-6  Convolution diagrams illustrating that many—but not all—out-of-pass-band signals, not related to any sample clock, do not correlate—Nyquist Zone 1 case.  The black lines in (c) indicate that the pass-band, as well as one out-of-band spectral signal (marked, and showing region where signals will correlate), which does correlate.*

Figure 3-7 (on this and the following 2 pages) illustrates the specific case of a "relaxed" anti-aliasing analog filter before digitization with SCFO sampling, Nyquist Zone 2. It shows that analog aliasing does not correlate, only aliasing due to the final frequency shift before correlation does—not an issue since such aliasing is absorbed within the discarded guard band anyway.

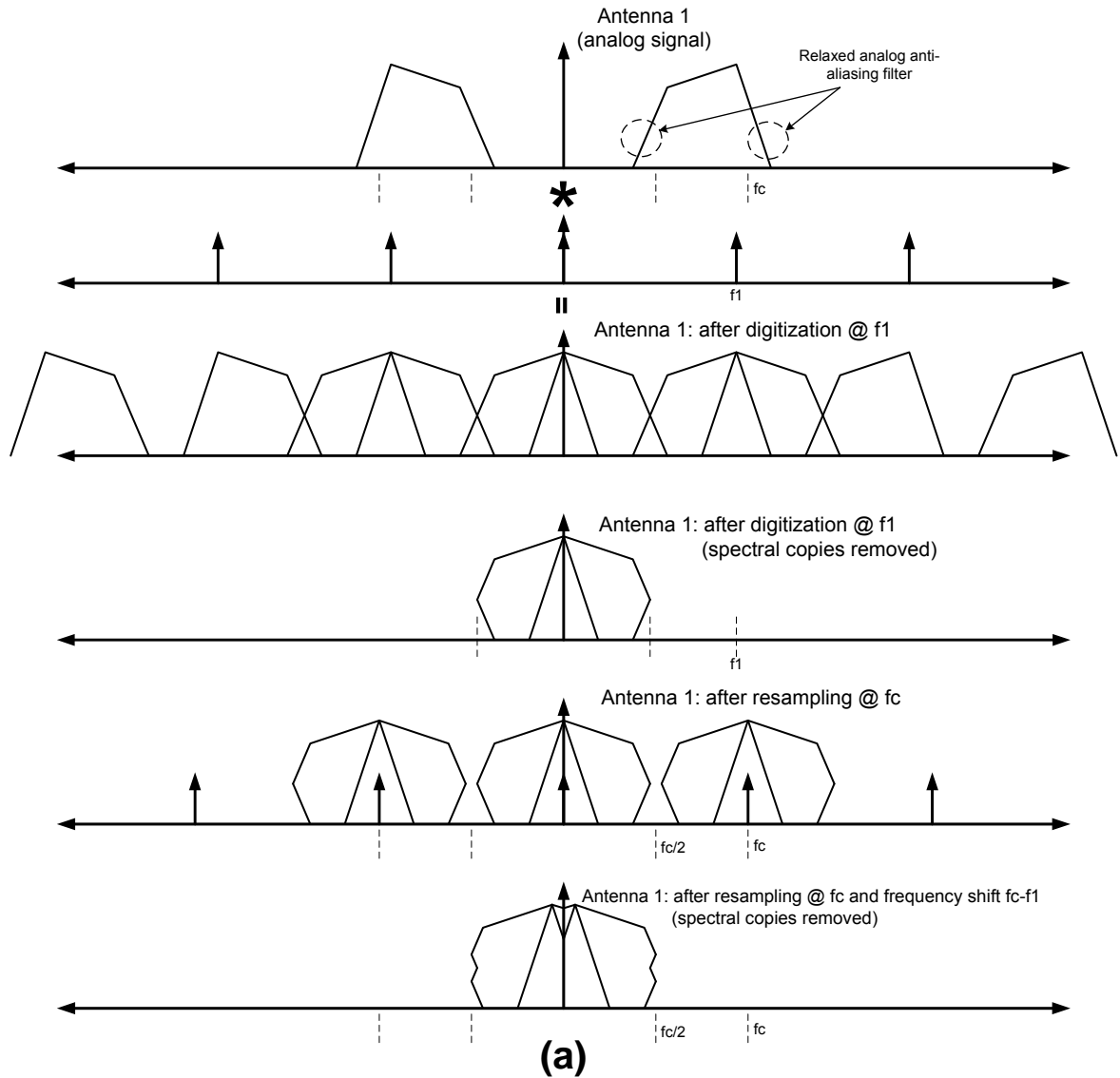

(a)

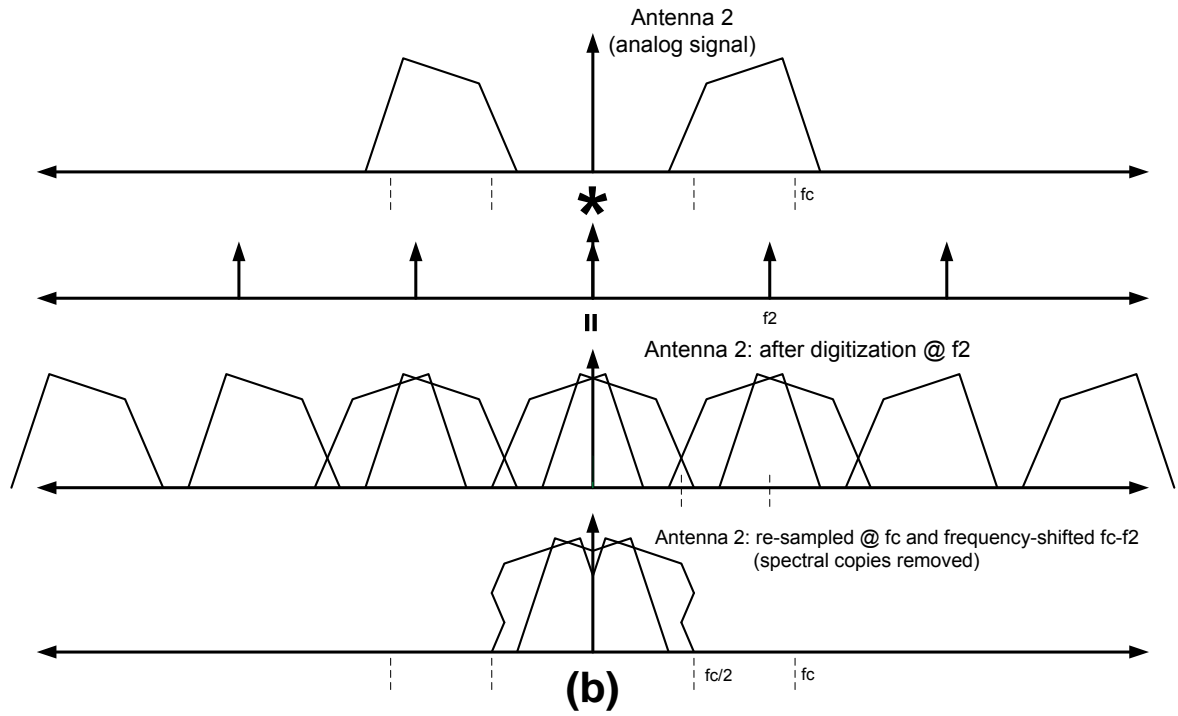

(b)

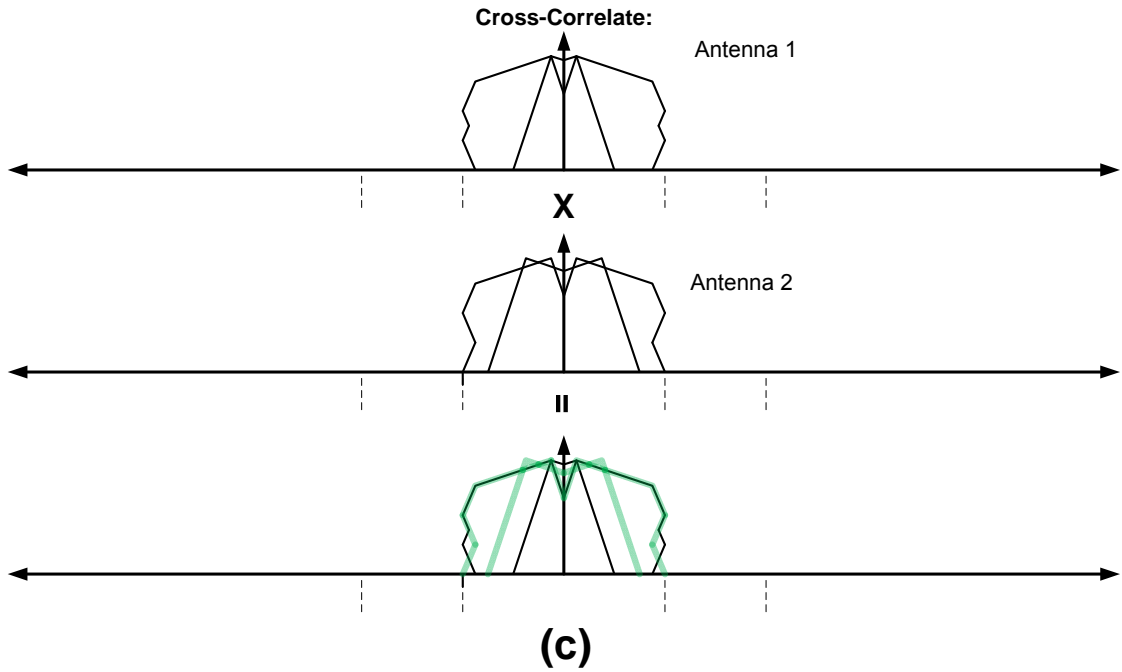

(c)

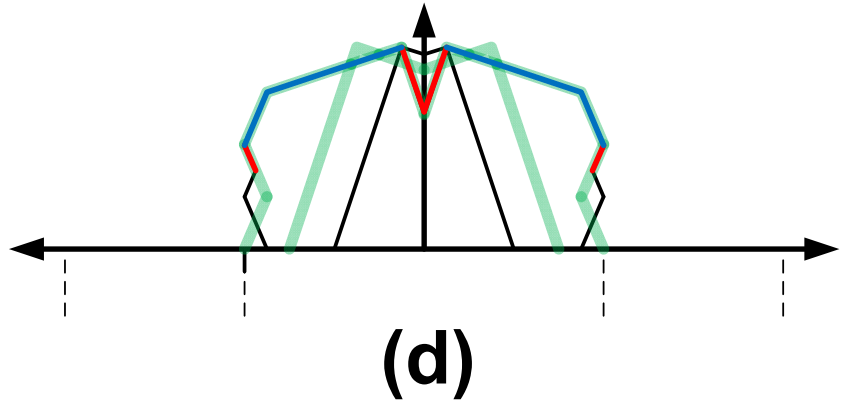

**(d)**

*Figure 3-7  Convolution diagrams for relaxed analog anti-aliasing filter, Nyquist Zone 2.  (a) is Antenna 1 sampling at frequency f1; (b) is Antenna 2 sampling at frequency f2; (c) is the cross-correlation, and (d) is an enlarged view showing (blue) the portion of the passband correlating, and (red) aliased correlation due to aliasing from the final frequency shift normally within the guard band anyway.  Relaxed analog anti-aliasing filter aliased portions do not correlate.*

This same effect of de-correlation of aliasing from a relaxed analog anti-aliasing filter occurs for Nyquist Zone 1 sampling as well (although not shown here), since the aliasing frequencies are a function of the sample frequency.

These facts mean that another advantageous side-effect of the SCFO scheme is that analog (or digital, as the case may be) anti-aliasing filters (prior to final common sample clock re-sampling) can have reduced performance where any spectral confusion/contamination is heavily suppressed and the effect of aliasing—and the design of the anti-aliasing filter—is primarily a SNR degradation consideration in the region where aliasing occurs, rather than a spectral contamination consideration.

## 4 SADT Implementation

At the moment, SADT is proceeding with two independent designs for the delivery of the sampler clock. One is by the Tsinghua University (THU), the other by the University of Western Australia (UWA).

The THU system transmits an optical carrier that is modulated with a 2 GHz signal derived from the central frequency reference. In the dish pedestal, the system transmits a 1 GHz modulated optical signal on a different optical wavelength, which travels to the CPF and is reflected back to the dish it came from. This measures the round trip phase variation. The variation in phase for the 2GHz one-way signal is the same as that for the 1GHz round-trip signal. The 1GHz phase variation is therefore used to correct the 2GHz incoming phase, resulting in a phase stable 1 GHz signal at the dish pedestal.

The 1 GHz signal is used to drive a synthesizer which generates the sampling clock for the Dish digitizer. The sampling clock depends on the band in question.

- Band 1: 4.0 GHz
- Band 2: 4.0 GHz
- Band 3: 3.2 GHz
- Band 4: 5.4 GHz
- Band 5: TBD.

For band 5, there is not yet an agreed upon value. The sampling frequency for band 5 was originally agreed to be 32 GHz, but might be changed to 30 GHz. The clock that SADT will provide would be a sub-harmonic of this, e.g. 6 GHz (5x) or 7.5 GHz (4x).

Generating each of the above frequencies out of the stabilized 1 GHz signal is straightforward. To add the capability of generating an offset sampling clock requires a much more complicated synthesizer. Such a synthesizer employs a DDS to generate the small offset frequencies, and includes an interface for SAT.LMC to not only select the band, but also the output frequency of the oscillator.

The UWA system operates in a fundamentally different fashion. The reference frequency in this case is transported as the frequency difference between two optical signals that are generated from a single laser.

The output of this laser is equally divided in two signals. Each of these optical signals has its frequency shifted by a particular amount, such that the difference between both frequencies is fixed at 8.0 GHz, independent of the exact frequency of the 1550nm laser. The two signals are combined again and transmitted over a single fiber.

At the Dish location, both signals are mixed together in a photo-detector, generating a 8 GHz beatnote, from which the required sampling clock is derived. A percentage of the signal is also reflected back, picking up a known frequency shift after passing through a set of acousto-optical modulators and, after mixing with a reference optical signal, produces an 80 MHz electronic signal that contains a measure of the phase fluctuations of the fibre link. The servo system in the Transmitter Module in the CPF is used to compensate these fluctuations, resulting in a stabilized signal at the Dish.

The SCFO is generated independently for each optical link, by using SAT.LMC to program each Transmitter Module's DDS to apply a unique frequency offset to the servos' 80 MHz local oscillator signals. This DDS is required in the design even without the SCFO scheme, and therefore the SCFO scheme has no cost impact for this design.

Separately from the frequency distribution systems above, SADT will also employ a system for the distribution of absolute time through the telescopes. This fixes the received signals to the local realization of UTC, and from that to actual UTC. The main scientific driver for this is the timing of pulsar signals, but it is also required for VLBI and establishing timestamp and delay model application time accuracy in the Mid.CBF.  Absolute time distribution is done by an already existing COTS system called 'White Rabbit'.

The White Rabbit system also uses fiber optics to distribute a clock signal across each telescope.  It works by measuring the round trip time on a single fiber to high accuracy, and calculating the delay incurred on the one-way trip from the sender to receiver. The system them compensates for this delay so that its PPS output is closely aligned with that of the central SKA clock ensemble, despite fiber delays of up more than 1ms on the longest paths.

The frequency distribution and absolute time distribution are completely independent. The frequency distribution has very high stability to ensure coherence across the array, but does not carry any time information.  The time distribution system cannot achieve the same stability, and therefore both systems cannot be expected to be phase coherent.  Operationally, the telescope can be phased up using the absolute time signal, and then maintain coherence through the distributed reference clock.

However, an additional requirement is that the PPS signal that is provided by SADT must be kept synchronous to the reference clock. This can be achieved by the use of a 'synchronizer', which re-samples the PPS signal in the clock domain of the sampling clock.  The design of such a synchronizer is significantly complicated[4] if the sampling clock does not have fixed value.

SADT has been required by the SKA Office to assume in its designs that the SCFO system will be employed in SKA1-Mid.  Both frequency distribution designs and the submitted SADT costings already include provisions for the SCFO system.  However, this assumption was only stated in the vaguest terms, and did not include sufficient detail (such as the required frequency step size and tuning range), so a complete design can only be started after the SCFO decision has been formally made and its requirements known to us.

Recent discussion between the 3 authors of this report (representing CSP, SADT, and DISH) indicate that the assumptions SADT made about frequency resolution and in particular regarding the impacts of non-integer Hz resolution using a DDS can be handled (as described in section 2.2) by DISH and CSP if known apriori.

For SADT to move forward into detailed design and prototyping, the following assumptions should be made regarding the SCFO scheme:

- Non-integer Hz frequency resolution is acceptable as long as the sample clock *offset* frequency can be defined precisely as a rational number.

- Maximum ~100 Hz final sample clock tuning resolution with no exact/special frequency resolution required.  For Band 5 where there is more guard band bandwidth, up to ~1 kHz final sample clock tuning resolution is acceptable.  These resolutions allow flexibility in offset frequency choice if, for example, some future requirement/effect requires use of specially-selected frequencies (e.g. prime numbers).

- Maximum frequency offset for any antenna ~+/-1 MHz required, ~+/-10 MHz desired[5].

- The absolute phase of any antenna sample clock at the beginning of an observe block (see section 2.2 for a definition of "observe block") is undefined and established with a sky calibration observation.

---

[4] Author Carlson note: for the case of distribution of a clock that cannot be easily managed by a complex logic device such as an FPGA, ~≤ 500 MHz.
[5] For SKA1 to allow for even more self-interference suppression and for a design amenable to SKA2 with at least 10 kHz offsets on 2000 antennas.

- The time between White Rabbit-delivered 1PPS pulses at the antenna is, on average exactly 1.0 seconds.  Generally, then, the 1PPS and the offset sample clock (fa) will be drifting relative to each other.  As mentioned above, a digital synchronization circuit is required to sample the 1PPS into the fa clock domain for further use.

- If, during an observe block, an antenna goes down and then comes back up, it is accepted that its phase must be re-established with the next sky calibration observation[6] (before visibilities on any of its baselines can be used for science).  This is the case anyway without the SCFO scheme since the White Rabbit-delivered 1PPS is of unknown phase relative to the sampler clock—an impact that means that the CSP Mid.CBF will only "loosely" use the antenna 1PPS for timing alignment rather than precisely and predictably.

- It must be possible to turn off SCFO's and use the same sample clock frequency in every antenna.  This provides a fallback position if some future unforeseen effect of the method occurs for some science observations.

---

[6] For bands 3, 4, and 5.  For bands 1 and 2 it is likely that there will always be an in-beam calibrator.

## 5 DISH Implementation

Under the currently agreed scheme for sampling clock delivery to DISH, the SPFRx (DISH Digitizer) receives from the SADT.SAT a fully conditioned and ready to use clock at a frequency required by the selected band[7]. Maintaining the exacting timing of the clock pulses is a sole SADT responsibility which should include applying the offsets. Consequently this would make offsets completely transparent to the SPFRx.

As agreed in recent discussions between the authors of this report, a synchronizer circuit is required in DISH for sampling the 1PPS, where a varying number of clock cycles between White Rabbit-delivered 1PPS pulses as well as 1PPS pulses drifting relative to the sample clock are accepted if the SCFO scheme is employed. This synchronizer circuit generally samples the 1PPS on multiple incremental clock phases and then chooses the phase in the center of the clock phases showing the pulse is present. Internal logic then re-samples the pulse to the original (0) clock phase thereby guaranteeing it is properly and unambiguously sampled. As the 1PPS slowly drifts around with time this circuit guarantees it is always properly sampled. The clock phase that was used for sampling the pulse can be passed on to CSP to provide additional information on 1PPS pulse drift and absolute White Rabbit performance.

An additional complication arises in the context of Band 5, which is sampled as a whole at 30 GSps and then filtered and re-sampled by polyphase filters at 6 GSps to extract science sub-bands 5A and 5B of a 2.5 GHz bandwidth (Figure 5-1).

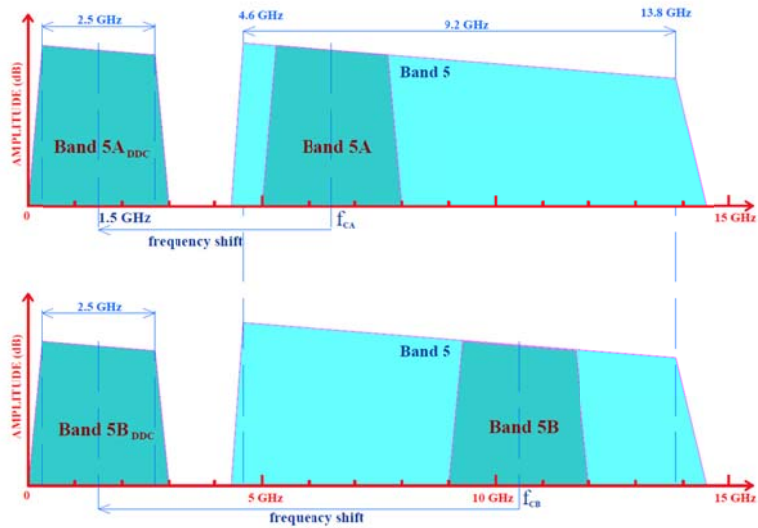

*Figure 5-1 Digital down conversion scheme for extracting sub-bands 5A and 5B into 3GHz baseband*

It is these sub-bands that are passed on to the CSP. To have the same offset amount as in Bands 1-4, the offsets five times higher must be applied to the 30 GHz clock. From the SPFRx perspective it may be advantageous that the SADT.SAT delivers SPFRx a 6 GHz clock which is internally multiplied to 30 GHz by the SPFRx.

---

[7] This may change if the 1PPS synchronizer circuit mentioned below is problematic at the full sample clock frequencies, TBC.

# 6 CSP Implementation

## 6.1 Simplified Equivalent Case

The analog signal at each antenna is originally sampled (digitized) at a sample rate unique to each antenna, $f_a$ and then re-sampled to a common sample rate $f_c$ before channelization and correlation, providing the benefits as demonstrated in section 3. Re-sampling could occur anywhere prior to further processing in the CSP, but it makes sense for this re-sampling operation to occur in the CSP since a) it is a complex operation using some power and, b) it is important to avoid any possibility of contamination of the antenna analog signal with the common sample clock $f_c$.

In the simplest terms, re-sampling to a common clock frequency $f_c$ is equivalent to feeding the original antenna digitized signal into an ideal Digital to Analog Converter (DAC) and then into an Analog to Digital Converter (ADC) using sample clock $f_c$. This process is shown in Figure 6-1:

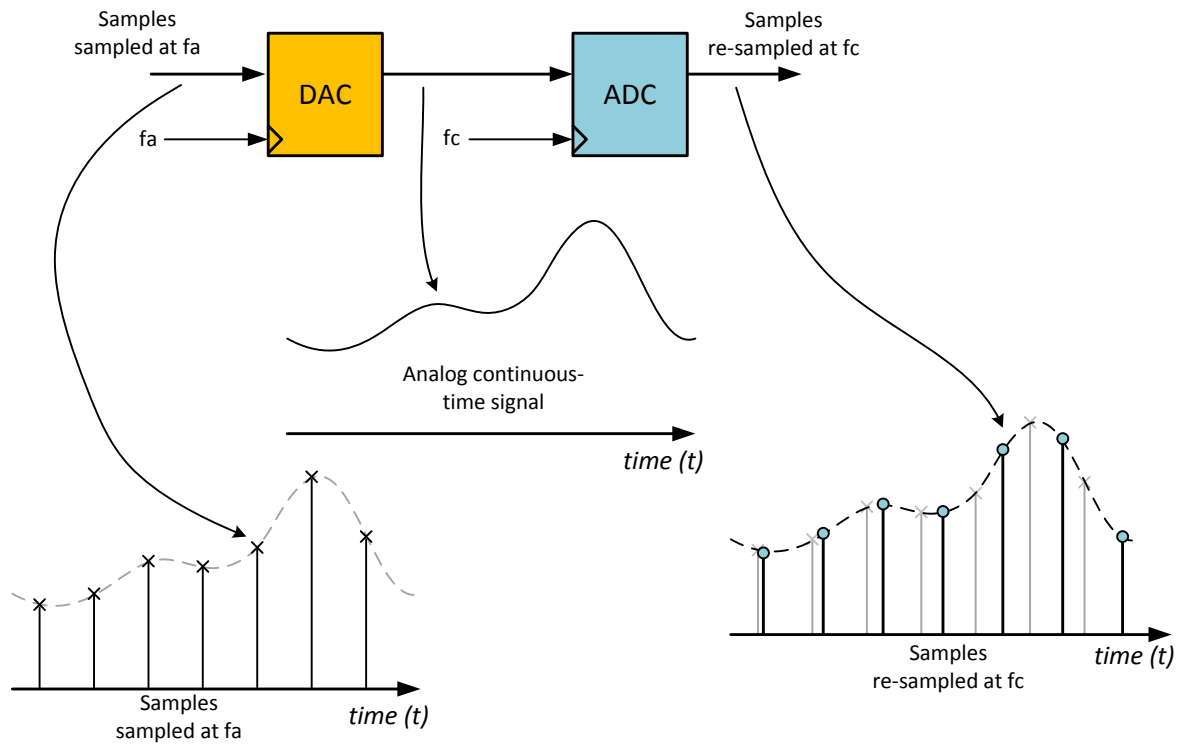

*Figure 6-1  Equivalent re-sampling circuit.  Samples sampled at $f_a$ are converted to an analog signal and then re-sampled by an ADC at sample rate $f_c$.  In this example, there is no amplitude quantization.*

Fundamentally in this process, if $f_c \neq f_a$, then the re-sampled signal will be over or under-sampled, potentially resulting in aliasing. However, in a practical system with guard bands and where $f_c$-$f_a$ is within some small fraction of the guard band, aliasing is not an issue. As well, examination of the figures in section 3 show that aliasing in the original sampled signal doesn't correlate anyway so any aliasing artefacts are highly attenuated[8].

Provided the re-sampling process is performed with sufficient care and noting the Nyquist sampling theorem, the re-sampled signal exactly retains and represents the original sampled signal with no fundamental artefacts of concern.

---

[8] A side-effect that can be used to advantage: analog anti-aliasing filters' performance can be relaxed since aliasing results in additive noise rather than spectral contamination—see section 3.

*Digital* re-sampling takes the same basic approach as above, except that a digital FIR interpolation filter is used to obtain the re-sampled data points, rather than the unwieldy and error-prone (due to unpredictable analog effects in a real circuit) process shown in Figure 6-1.

A simplified time-domain representation of the digital re-sampling process is shown in Figure 6-2, studied in some detail in [1], and with further results in section 7.

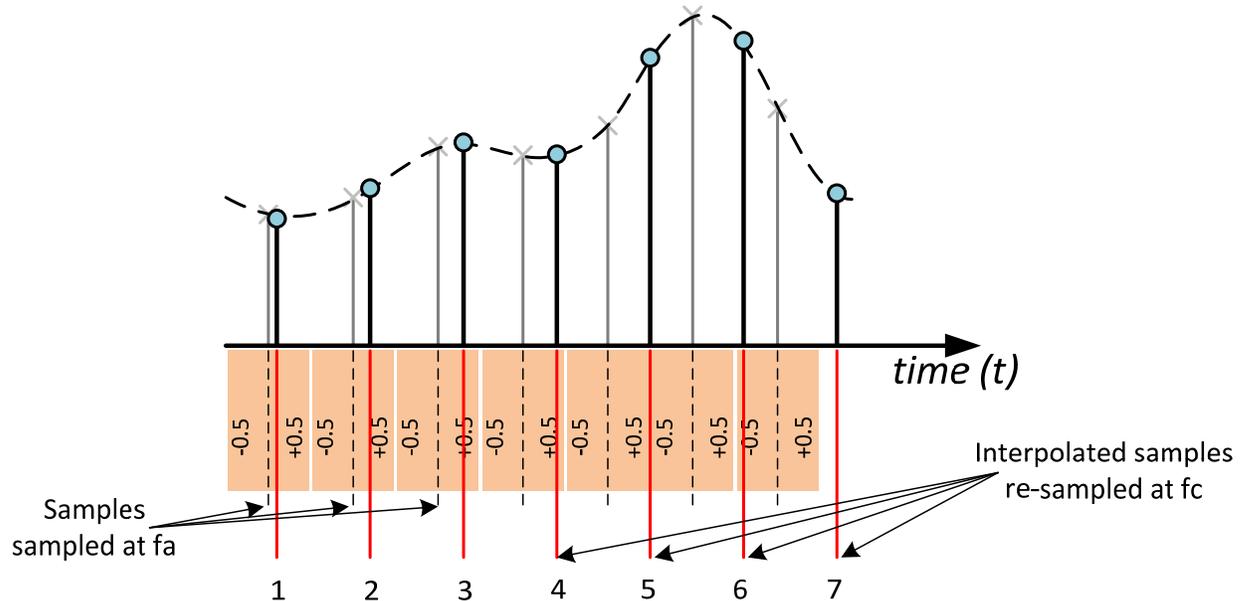

*Figure 6-2  The digital re-sampling process.  The original $f_a$-sampled digital samples (X) are passed through a +/-0.5 sample interpolation filter clocked at frequency $f_c$ to yield <u>different</u> interpolated values (O) at every $f_c$ sample time.  In this example, $f_c<f_a$, which requires skipping an input sample every so often, in this example at sample 5 when the interpolator is at +0.5 sample where subsequently at sample 6 the interpolator flips to (slightly less than) -0.5 sample.*

How well a realizable digital re-sampling interpolator works at the sample rates required by CSP Mid.CBF is of direct concern regarding the feasibility of implementing the SCFO scheme in the Mid telescope.  The rest of this section is devoted to exploring implementation issues in sufficient detail to judge feasibility.

## 6.2   Antenna $f_a$ and Re-sampler $f_c$ ≤ Digital Clock Rate

For the case where $f_a$ and $f_c$ are at sample rates ≤ a clock frequency that can be directly used in the digital processing hardware, the implementation of the re-sampler is relatively straightforward.  A block diagram of such an implementation is shown in Figure 6-3.  The length of the FIR filter (number of taps) and number of sets of tap coefficients between +/-0.5 samples (i.e. the size of the "Tap coef LUT") are explored in [1] and section 7 in further detail.

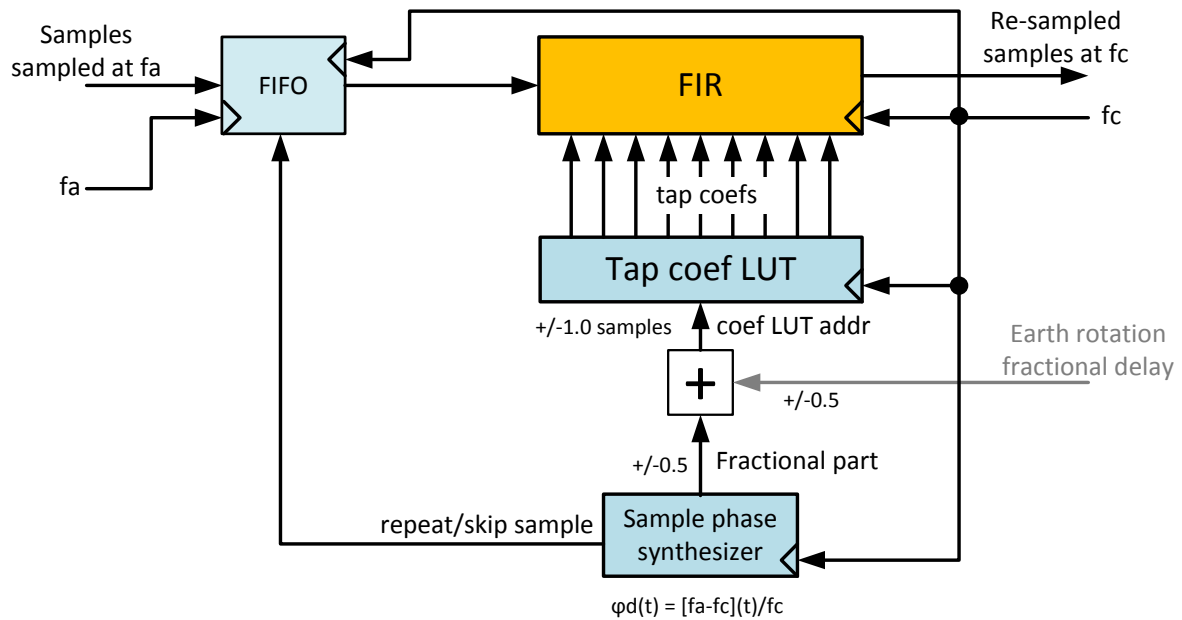

*Figure 6-3  Block diagram of the re-sampler where $f_a$ and $f_c$ are at sample rates that can be directly processed with digital hardware.  For current and next generation FPGAs this is 500 MHz to ~800 MHz (or even up to 1 GHz).  "Earth rotation fractional delay" is optional and is being considered by Mid.CBF for implementing wideband very fine (sub-sample) delay since all of the interpolation machinery is in place for it.*

The following points are of note regarding Figure 6-3:

1. The "Sample phase synthesizer", on every $f_c$ clock cycle, determines the interpolation filter fraction "delay phase" between +/-0.5 samples of delay.  It effectively synthesizes the difference frequency between $f_a$ and $f_c$.

2. The output of the phase synthesizer forms an address into the "Tap coef LUT" to select, *for that particular $f_c$ clock cycle*, the set of interpolation tap coefficients that are to be used to produce the FIR output sample.  This synthesizer also has an integer part that tells the FIFO if it is to repeat or skip a sample on that $f_c$ clock cycle.  A sample is occasionally repeated into the FIR if $f_a < f_c$ and skipped if $f_a > f_c$.  There is a momentary edge effect when a sample is repeated or skipped (see section 7) but since $(f_a-f_c)/f_c$ is normally very small, it is a very rare occurrence relative to the number of samples being processed, lasting only for the impulse response time of the filter.  For example, for a 2000 antenna array at a 1 kHz minimum frequency offset[9], the worst-case frequency offset is 2000 x 1 kHz = +/-1 MHz.  For a 1 GHz sample rate, 1 MHz / 1 GHz is approximately 0.1% of the time that a sample is skipped or repeated and since the impulse response of the filter > 10% amplitude is only a few taps, only ~0.3% of the data is impacted in any way.  Section 7 shows that the impact on the data every time there is a skip and repeat sample is of no consequence anyway.

3. The FIFO is kept centered at all times so that there is a constant delay between the input to the FIFO and the output of the FIR, skipping or repeating samples as demanded by the synthesizer.  Input and output data and address synchronization occurs since in reality the input clock is $f_c$ and samples are only clocked into the FIFO when discrete packetized samples are available.

---

[9] Providing ~38 dB spectral artefact attenuation in a 1 second integration time (~28 dB in 0.1 sec).

4. There is a delay between the selection of the particular set of tap coefficients and the resulting output of the FIR. This is constant and is therefore not an issue.

5. Handling of the 1PPS signal is not shown in Figure 6-3. It can easily be handled (synchronized to $f_c$) since it is already sampled and inserted into the samples streaming from the antenna. At the CSP end it is put through an equivalent delay of the FIFO and FIR, clocked at $f_c$.

In practise, the DISH digitizer clock rate $f_a$ is normally quite a bit larger than a clock rate that can be used directly in digital logic. It is therefore necessary to process data in Mid.CBF in a time de-multiplexed fashion, complicating the circuit of Figure 6-3 somewhat, but still with equivalent results. It is this topic that is covered in the next sub-section.

## 6.3 Antenna $f_a$ and Re-sampler $f_c$ > Digital Clock Rate

Under the conditions of $f_a$ and $f_c$ > the digital clock rate, time-de-multiplexed processing techniques must be used. For example, in Band 5, the sample clock rate $f_a$ is nominally 6 GHz—a frequency impossible to use directly in any reasonable digital logic implementation.

This problem is tackled by first developing a direct-form FIR implementation example operating at the actual/original sampling frequency we'll call $f_o$ and then mapping this to an example de-multiplexed structure and implementation operating at the de-multiplexed frequency $f_o/k$, where k is the de-multiplex factor. Figures and associated descriptions on the following few pages present the results of this approach.

Figure 6-4 is a direct form 9-tap FIR implementation with constant tap coefficients and with time slices and colours/labels set for a de-multiplexed factor of k=3 implementation. Samples are labelled as noted, with samples older than the reference "S0" negative (e.g. "S-8") and newer samples than the reference positive (e.g. "S2"). Some points of note in the figure are:

- There are 3 time slices, "a", "b", and "c" also coloured red, blue, and green for ease of identification. These are slices within each de-multiplexed clock (i.e. $f_o/3$) cycle.

- The time axis and associated samples above the FIR show the progression of samples in the shift register with time—one row for each $f_o$ digitizer clock cycle. For example, at time "t0a", samples in the FIR shift register starting at tap 0 are "S-6, S-7, S-8, etc.. One clock cycle later at "t0b" they are "S -5, S-6, S-7, etc. It will turn out that "t0" is the first de-multiplexed clock ($f_o/3$) cycle, "t1" is the second cycle, "t2" the third cycle etc.

- The output of the FIR, ignoring delays through the tap coefficient multipliers and adders are also noted for each $f_o$ time slice.

The direct-form FIR, mapped to a k=3 de-multiplexed structure, with identical sample and time slice marking is shown in Figure 6-5. Inputs and outputs are also annotated with simplified sample numbers (e.g. 0, 1, 2, 3, …), which are handy for the following figures' marking. Comparison of Figure 6-4 and Figure 6-5 reveals that they perform the identical mathematical operations noting that in Figure 6-5, 3 identical poly-phase FIR structures are required, each has 9 taps, each processes one time slice, and each tap operates at a clock frequency of $f_o/3$—a frequency amenable to implementation in digital logic.

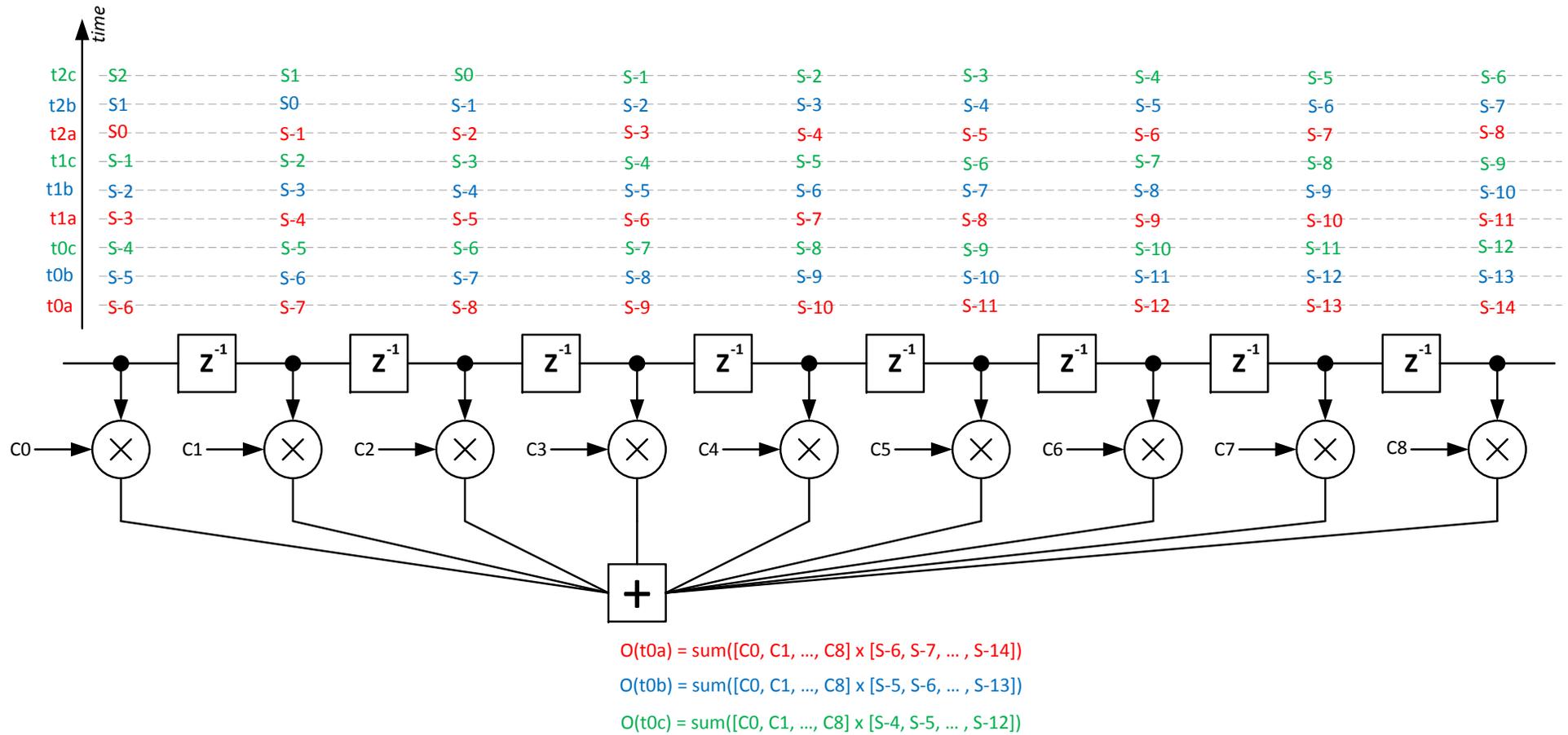

Figure 6-4  Direct form FIR implementation with constant tap coefficients at the digitizer sample rate but with colours and timeslices for a de-multiplexing factor of 3.  Samples ("Sn") relative to an arbitrary reference (S0) are noted with samples older than S0 negative (e.g. "S-14") and newer than S0 positive (e.g. "S2").

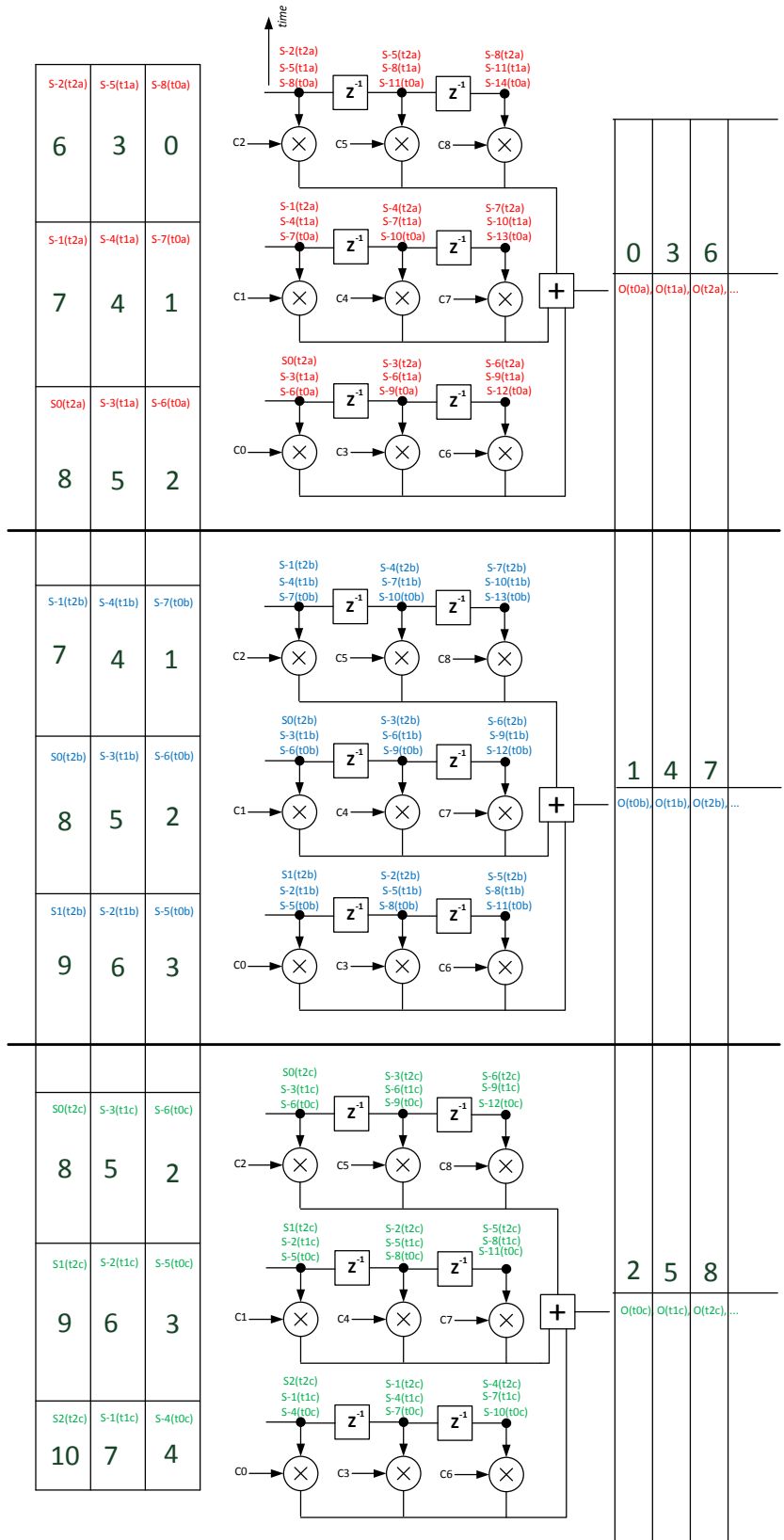

*Figure 6-5  Direct-form FIR of Figure 6-4 mapped to a k=3 de-multiplexed structure (annotated with simplified sample markings in large numbers.)*

The example of Figure 6-5 shows a classic trade-off of logic vs performance. In this case, 3X the logic running at 1/3$^{rd}$ the performance achieves the desired full performance result.

This example can be extended to any number of taps and any k de-multiplexing factor where the number of taps is normally an integer multiple of k. The actual value of k used in any implementation depends on logic performance and the original sample frequency $f_o$. For example, for Mid telescope Band 5 with $f_o$=6 GHz, with k=8, FIR logic operates at 750 MHz, very likely possible with the Altera Stratix-10 technology planned for Mid.CBF. For a 56-tap implementation, 56 x 8 = 448 actual taps per sampled data stream are required.

A simplified/collapsed representation of Figure 6-5 is shown in Figure 6-6, showing the input sample sequences, de-multiplexing commutator, and output samples. The input commutator must "barrel-roll" the samples across the poly-phase FIR blocks, however the outputs are already in the familiar time-de-multiplexed time sequence ready for further use.

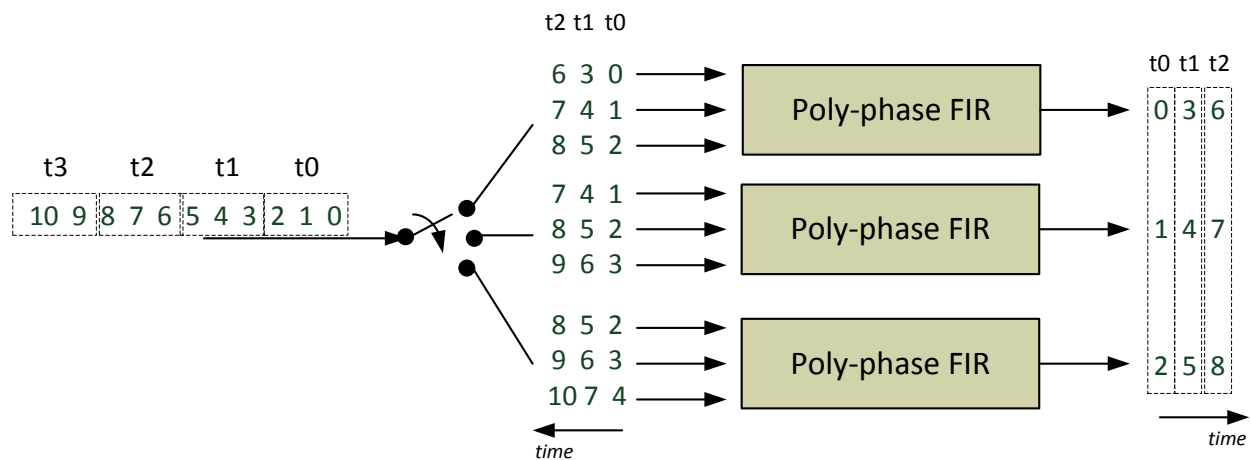

*Figure 6-6 Simplified/collapsed representation of Figure 6-5, for k=3 de-multiplexed FIR structure.*

A simplified combination of Figure 6-3 and Figure 6-6 is shown in Figure 6-7, not including the sample repeat/skip FIFO. This is the de-multiplexed re-sampling circuit block diagram.

The entire circuit operates at $f_o$/3 (renamed $f_c$/3 in the figure since it is re-sampling to the common clock frequency $f_c$) and optionally incorporates Earth rotation fractional sample delay.

A key aspect of this circuit is that each of the 3 poly-phase FIRs is identical in implementation and processes a unique time slice. This means that the sample interpolation phase coming from the "Sample phase synthesizer" can be calculated for each $f_c$ time slice of the $f_c$/3 cycle and applied to that poly-phase FIR uniquely and independently. The time slices and Sample phase synthesizer values for each time slice are shown in Figure 6-8, noting that in reality there is only one $f_c$/3 clock and that all 3 sample phases are generated and used at the same time.

The other important implications are that a) each poly-phase FIR is identical with identical tap coefficient memory LUTs and, b) each poly-phase FIR needs its own set of coefficient memory LUTs since each FIR has a slightly different sample phase and therefore LUT address.

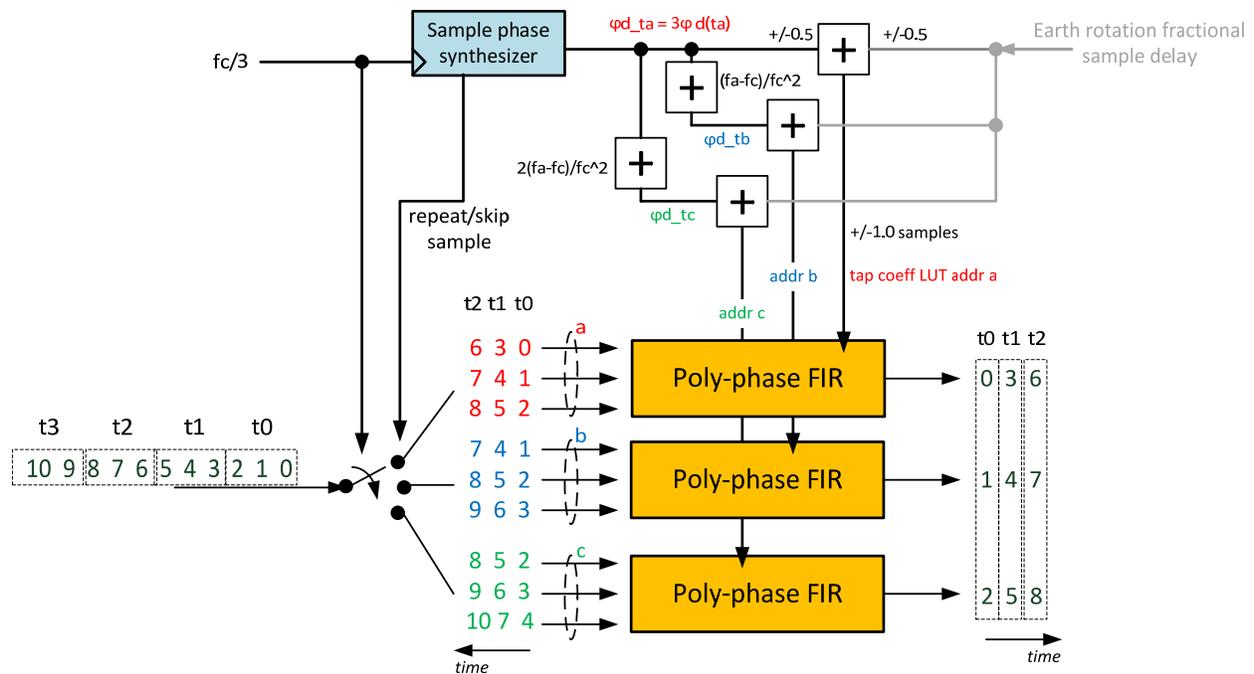

*Figure 6-7  Re-sampling de-multiplexed k=3 block diagram, formed by combining Figure 6-3 and Figure 6-6.*

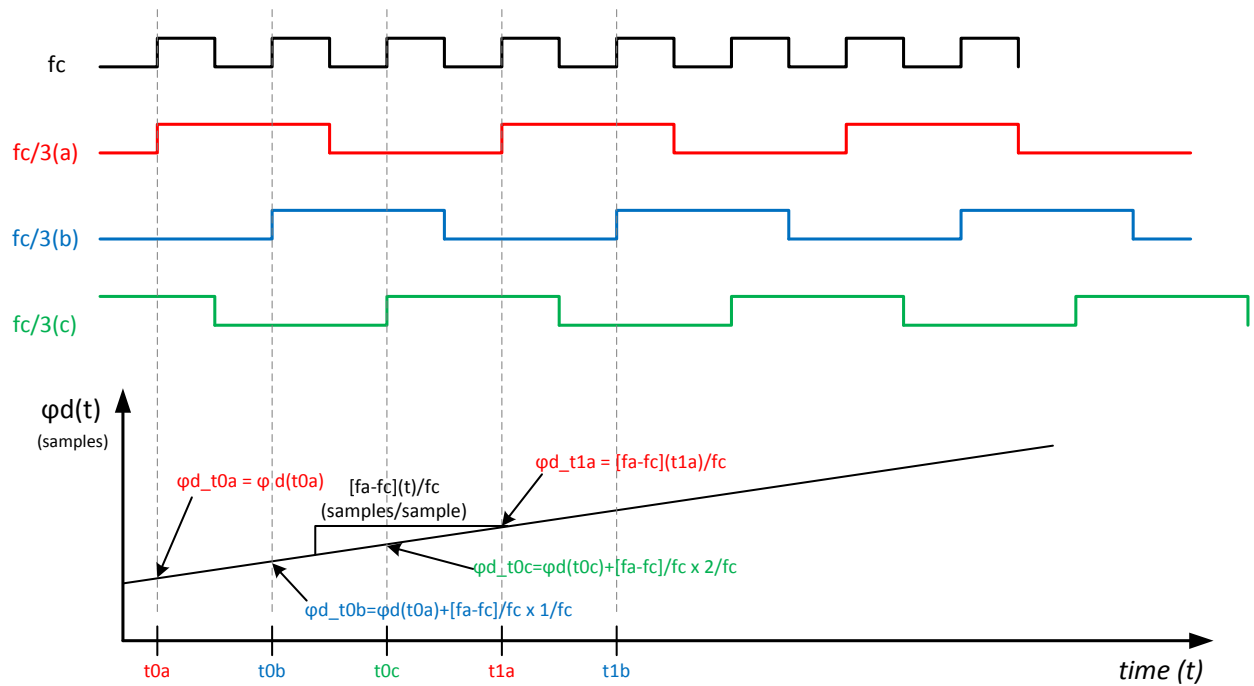

*Figure 6-8  Example k=3 sample phase synthesizer outputs showing the direct form clock $f_c$ and each of the $f_c/3$ clock phases. In reality in the circuit of Figure 6-7 there is only one $f_c/3$ clock and all 3 sample phases are generated at the same time since they are—for a linear sample phase synthesizer—predictable with the equations shown in this figure.*

As discussed extensively in section 3 re-sampling for anything other than Nyquist Zone 1 requires a final frequency shift of the difference between the antenna sample frequency $f_a$ and the resampling frequency $f_c$. This method requires a digital single-sideband (SSB) mixer, which itself contains a Hilbert-transform FIR filter. However, both re-sampling/interpolation and the SSB Hilbert transform FIR can be integrated into one structure, a simplified version shown in Figure 6-9:

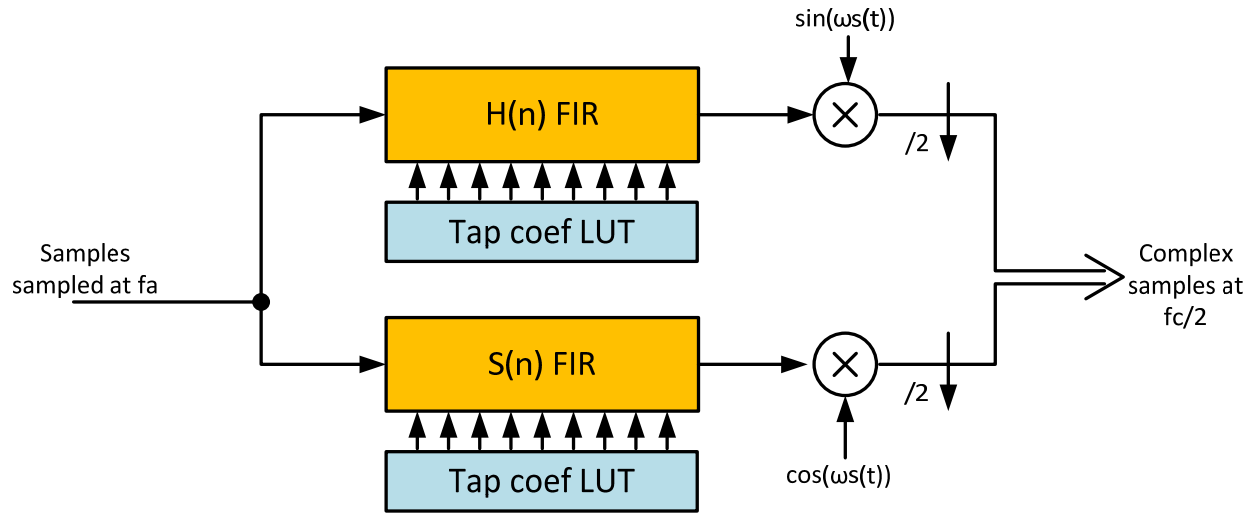

*Figure 6-9  Re-sampling circuit with integrated Hilbert transform and frequency shifter (SSB mixer). For a complex output with /2 decimation as shown, H(n) and S(n) are half-band interpolating filters. The Sample phase synthesizer is not shown.*

In Figure 6-9 a complex output is shown, which is convenient for down-stream channelizer use, however, a real output could be generated with this same structure by including a final adder after the sine and cosine multipliers. In a de-multiplexed structure implementation, the sine and cosine mixers require one actual mixer (sine/cosine LUT and multiplier) per de-multiplexed stream (3 as indicated in Figure 6-7 for k=3), with phases for each stream derived from one phase synthesizer (i.e. predictable/calculated phase offset for each de-multiplexed stream).

In Mid.CBF, the current plan is to integrate the earth-rotation fractional-sample very fine delay correction into the interpolating re-sampling filter as indicated in Figure 6-3 as well as incorporate the $f_c$-$f_a$ frequency shift, earth rotation fringe phase correction, frequency shift for complex signal generation and, if required, Doppler frequency shift correction to a required reference location (geo-center or bary-center). The latter Doppler correction only if such a correction can be used by imaging, PSS, and PST since the resulting wideband signal is used by all of those channelizers, although it is possible to have more than one mixer output section, each applying required phase rotations for downstream use (e.g. everything including Doppler correction for imaging, and everything except Doppler correction for PSS and/or PST).

Note that in Figure 6-9 the Tap coef LUTs for the in-phase (S(n)) FIR and the quadrature (H(n)) FIR can be shared, reducing LUT memory requirements by a factor of 2.

### 6.4   Subtle Synchronization Issues

As established in the previous SADT implementation section 4 the stream of samples and 1PPS marks coming from the antenna to the Mid.CBF is such that there are a varying number of samples between successive 1PPS marks. This is the case with or without the SCFO scheme active because the White

Rabbit 1PPS and the phase-compensated sample clock delivered by SADT are not phase synchronous in the antenna.

What this means at the CSP end is that the 1PPS mark from the antenna can be used only to establish approximate time alignment of the 1PPS delivered to the CSP from within the KAPB (the "KAPB 1PPS") to actual UTC, but can't be used for any other synchronization. This is done by, at the beginning of an observe block (see section 2.2 for definition of "observe block"), setting the delay model to 0 and then adjusting the depth of the delay tracking FIFO so that the antenna 1PPS fed through the FIFO, and the KAPB 1PPS delayed for all antennas by a constant ~½ FIFO depth and positioned at the output of the FIFO, line up in time (at the output of the FIFO.) The antenna 1PPS will "jump around" (relative to the KAPB 1PPS) by a few samples up to the accuracy that it is delivered to the antenna by SADT but, over a few seconds its time centroid can be lined up to the KAPB 1PPS, thereby absorbing all of the delay from the antenna to the CSP and "calibrating" the KAPB 1PPS as the reference. This is done for each antenna independently. If, subsequently, the fiber from the antenna to the CSP stretches and/or contracts, the wander is automatically absorbed in the delay tracking FIFO since everything is discrete time.

Ideally, the commutator circuit of Figure 6-7 is set to a known phase every 1PPS mark from the antenna, however, this is not possible for the reasons stated previously and the fact that for the SCFO scheme the sample clock $f_a$ is not in integer Hz. It may, however, be possible to use the 1PPS mark from within the KAPB for this purpose since it is exactly locked and synchronized to the 100 MHz clock provided to CSP from SADT, TBC. This may, however, be unnecessary or of little utility since the sample clock phase and 1PPS used in the antenna is fly-wheeling over the course of an observe block and so fly-wheeling the commutator circuit is no different. The consequence of this is that if the CSP "F-part" electronics for any particular antenna goes down and comes back up, phase for all baselines of that antenna will need a sky calibration to re-establish phase. Use of the KAPB 1PPS for synchronizing the commutator—which is deterministic—may avoid this problem. Further modelling and design work is required to determine if such synchronization is feasible, particularly when the antenna sample clock $f_a$ is not an integer 1 Hz multiple.

## 6.5 Logic Implementation

The worst-case performance-driven requirement of Mid.CBF for the re-sampler is Band 5. In Band 5, 4 x 6 Gs/s re-samplers must be implemented. Applying the de-multiplex structure of Figure 6-5, the complex implementation of Figure 6-9, 56 tap[10] FIRs, and k=8 de-multiplexing factor results in:

- Each complex FIR requires 56 x 8 x 2 = 896 tap multipliers. For 4 streams this is 3584 multipliers, each running at 6 GHz/8 = 750 MHz.

- Assuming sharing of Tap coef LUTs between the S(n) FIR and the H(n) FIR, 448 LUT memories are required. Assuming 1024 points between -0.5 and +0.5 samples[11] and 20-bit tap coefficient values out of the LUT, requires 1792, 1024 x 20-bit memories. This memory size is chosen because it exactly fits the Altera M20K memory block structure, provides the requisite 19-bit coefficient for the 18x19 multiplier, and has acceptable interpolation performance.

---

[10] As shown in section 7 and integer-divisible by k=8.
[11] Even though the incorporation of the very fine delay for earth rotation delay correction requires values between +/-1.0 samples, the integer part is incorporated into the commutator logic that barrel rolls the signals across the poly-phase FIRs.

- The k=8 de-multiplexed complex mixer requires 16 multipliers, each multiplier with a $1024^{12}$ x 20-bit (TBC) M20k memory for sine/cosine lookup. Therefore, another 4 x 16 = 64 multipliers and 64 M20k memory blocks.

Summing up:

Total multipliers = 3584 + 64 = 3648.

Total M20k memory blocks = 1792 + 64 = 1856.

The smallest Altera Stratix-10 device that is footprint compatible with the planned 96 SERDES 50 x 50 mm FPGA for implementing the re-sampler and some other functions [5] is the GX1650 [6]. This device has 6290 multipliers and 5851 M20k memory blocks. The Band 5 re-sampler, including very fine earth rotation delay, $f_c - f_a$ frequency shift, and earth-rotation fringe phase correction requires 3648/6290 ~= <u>58% of the multipliers</u> and 1856/5851 ~= <u>32% of the M20k memory blocks</u>.

The adder tree for a 48-tap FIR requires ~56 adders, each one ~37 bits, weighted average, for a total of ~2072 adder bits. For k=8, the number of adder bits is 8 x 2072 = 16576 and for 4 streams is 66304. For safety, double this for pipelining or ~130k LEs. Each adder bit requires one Logic Element (LE).

The FIR shift registers, assuming 8-bit words, require 56 taps x 8 bits x k=8 x 4 streams = 14336 LEs.

All other logic for the commutator and controller, delay phase synthesizers (64-bit point/slope=~256 per FIR), frequency shift+earth rotation phase synthesizer (64-bit point/slope), etc. might amount to ~10k LEs, maybe 20k LEs max.

Summing up:

Total LEs ~= 130k + 14k + 20k ~= 170k.

The 10SG165 devices has 1.624 MLEs, therefore the re-sampler logic utilization is ~170k/1.6M ~= <u>11% logic</u>.

Plugging all of these numbers into the Altera Stratix-10 power estimator spreadsheet (March 2016 version), 50% switching percentage, and 750 MHz yields a total power in the 20-50 W range[13]. This estimate doesn't include SERDES I/O power since such I/O is required anyway assuming a second FPGA is required for F-part processing.

Thus, the re-sampler for the SCFO scheme seems entirely reasonable as far as logic implementation and the *additional* power required to implement it is 197 x ~50 W ~= 10 kW. However, it really is not an additional power or logic cost since VLBI mode requires re-sampling to VLBI standard frequencies— either on the front-end with wideband re-samplers per beam or on the back-end with bandwidth synthesis and re-sampling. For VLBI mode, it may be the case that 56 taps per re-sampling FIR may not be possible—although the Mid.CBF F-Part architecture is still in flux and recent thinking is such that it may be that 2 VLBI beams requiring 2 wideband re-samplers may be implemented in each of 2 FPGAs, rather than 4 beams in one FPGA as indicated in [5] section 6.1 Figure 6-4. See [5] section 9 for further discussion on VLBI-mode implementation trade-offs.

---

[12] Still need to verify that 10-bit phase width into the sine/cosine LUTs is sufficient. If not, more memory is required.

[13] Exact numbers are NDA-protected and subject to change, but this is a fairly confident range.

# 7 Modelling

This section (all modelling results in this section provided by Thushara Gunaratne from NRC) includes the key modeling artefacts for re-sampling as well as containing some additional modeling results not included or covered in sufficient detail in [1].

Figure 7-1 is a short series of time-domain samples—including the original analog signal—illustrating that re-sampling interpolation accomplishes what is described in Figure 6-1 and Figure 6-2. In this example (as for all other re-sampling modelling) a broadband signal is emulated as a sum of sine functions with randomized amplitudes, frequencies, and phases to allow directly calculable comparisons between the original signal, the original sampled signal, and the re-sampled signal. As well, data is floating point but filter tap coefficients are 19-bit, with 1024 steps between +/-0.5 samples and 56 tap coefficients. Floating-point samples and re-samples are used to show how they track the analog signal—refer to Figure 7-6 for a similar time-domain plot with 4-bit initial sampling.

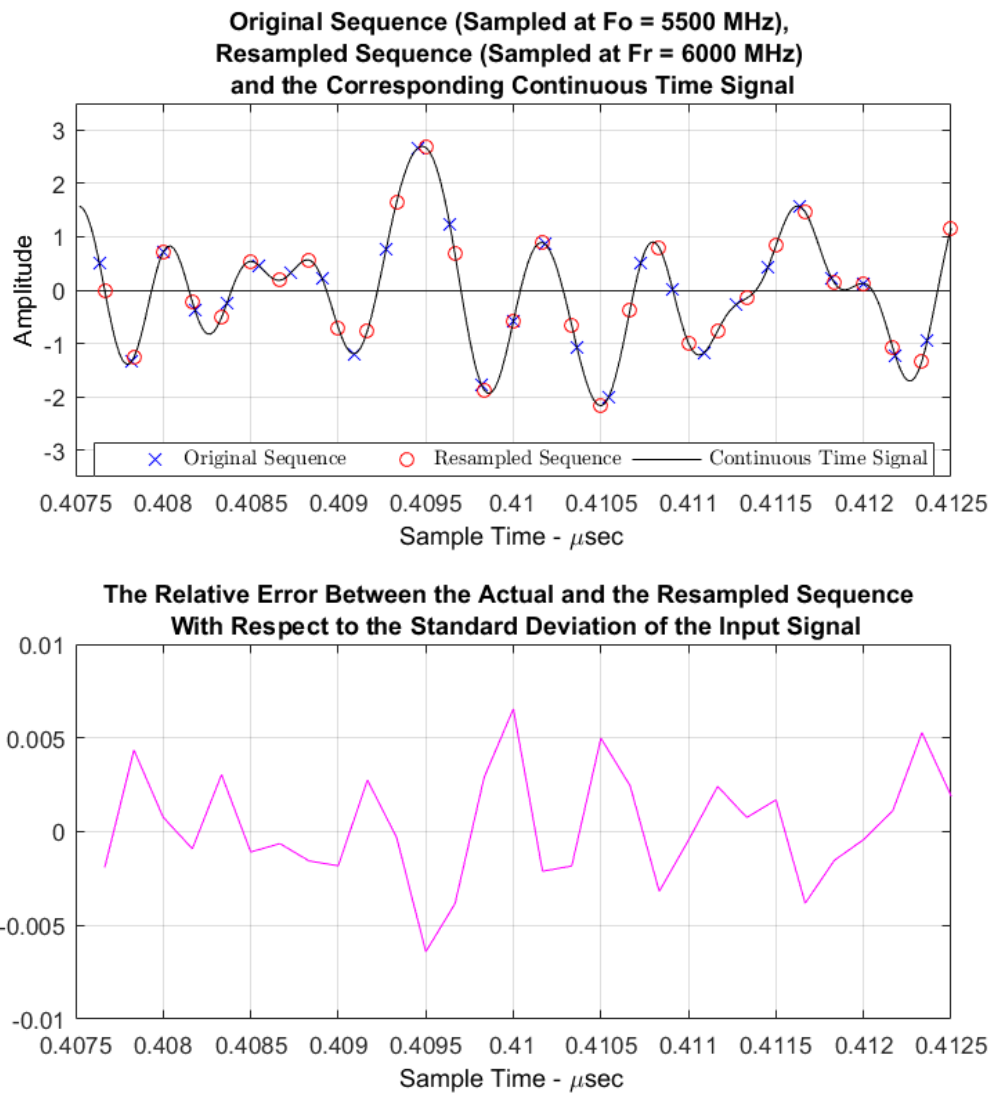

*Figure 7-1  Time-domain re-sampling example illustrating re-sampling as described in Figure 6-1 and Figure 6-2.*

The next 2 plots indicate the performance of the 56-tap re-sampling filter.

Figure 7-2 is the filter response, magnitude vs frequency and actual delay vs frequency both with the delay index of 0-1023 implementing the full range of +/-0.5 samples of delay indicated, for the case of floating-point tap coefficients. The bold black dashed lines in the upper plot indicate the spectrally pure region where filter performance is expected to be within limits to not be unduly inaccurate for the pass-band expected from DISH Band 5, i.e. starting at DC: 250 MHz guard band, 2.5 GHz pass-band, 250 MHz guard band.

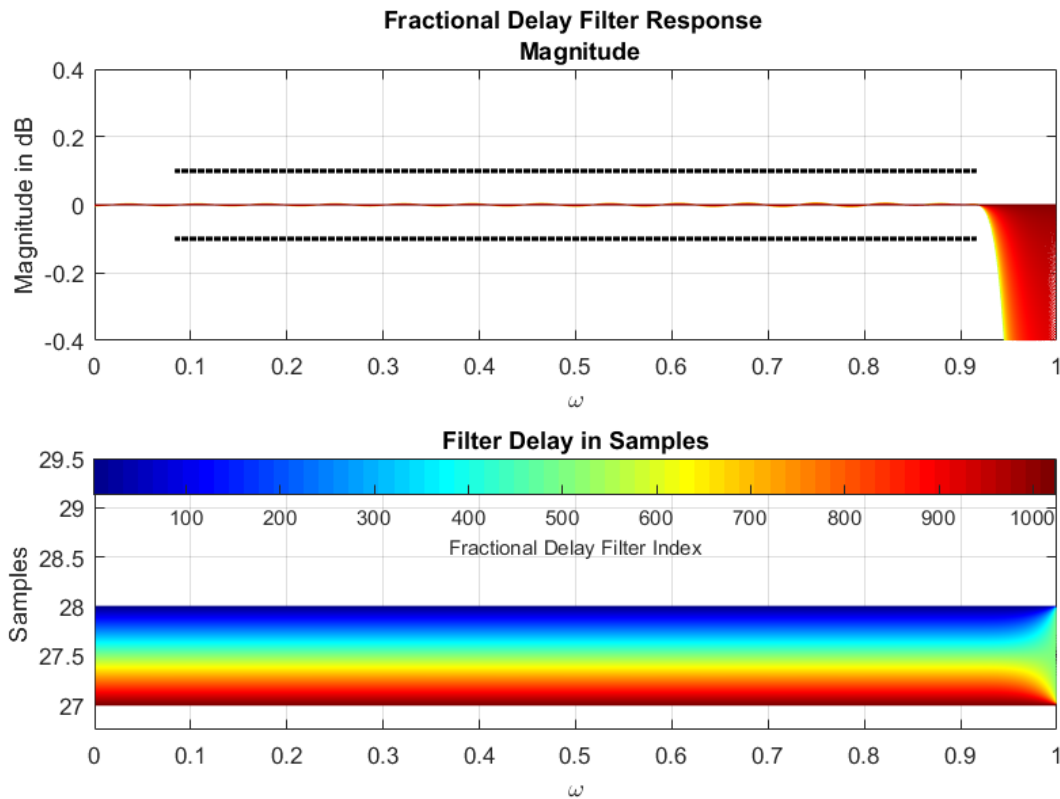

*Figure 7-2 56-tap filter performance with floating-point tap coefficients. The bold black dashed lines in the upper plot indicate the region of Band 5 pass-band.*

Figure 7-3 is a similar plot to the above figure except it is for 19-bit tap coefficients[14] showing the delay *error* in units of $10^{-4}$ samples with the required spectrally pure region for Band 5 indicated. Here we can see the effect of the sampling index changing from 0-1023 over +/-0.5 samples of delay. This will slightly modify the amplitude and delay of each sample—however with re-sampling between two different sample rates over any reasonably small correlator integration time, these effects will average out as the full range of +/-0.5 samples of interpolation delay sweeps out many times.

The calculated coherence loss at f~=0.875 due to ~0.2x$10^{-4}$ samples pk-pk of delay is 1 - sinc($\phi_{pp}$) [7] (for uniform phase distribution across the pk-pk swing), where $\phi_{pp}$~= π x 0.875 x 0.2x$10^{-4}$, calculated as a vanishingly small value. It is essentially the same across the required spectrally-pure band.

An additional frequency-dependent effect, illustrated in Figure 7-4, causes a correlated coherence loss vs frequency, but it is so small ($10^{-6}$) so as not to be a concern. This coherence loss can modulate the correlated amplitude when the re-sampling circuit is also used for very fine/fractional delay correction if the delay rate beats with the correlator integration time, but again it is a vanishingly small effect and very likely needs no post-correlation correction[15].

---

[14] The implementation in the FPGA, with its 18x19 multipliers.

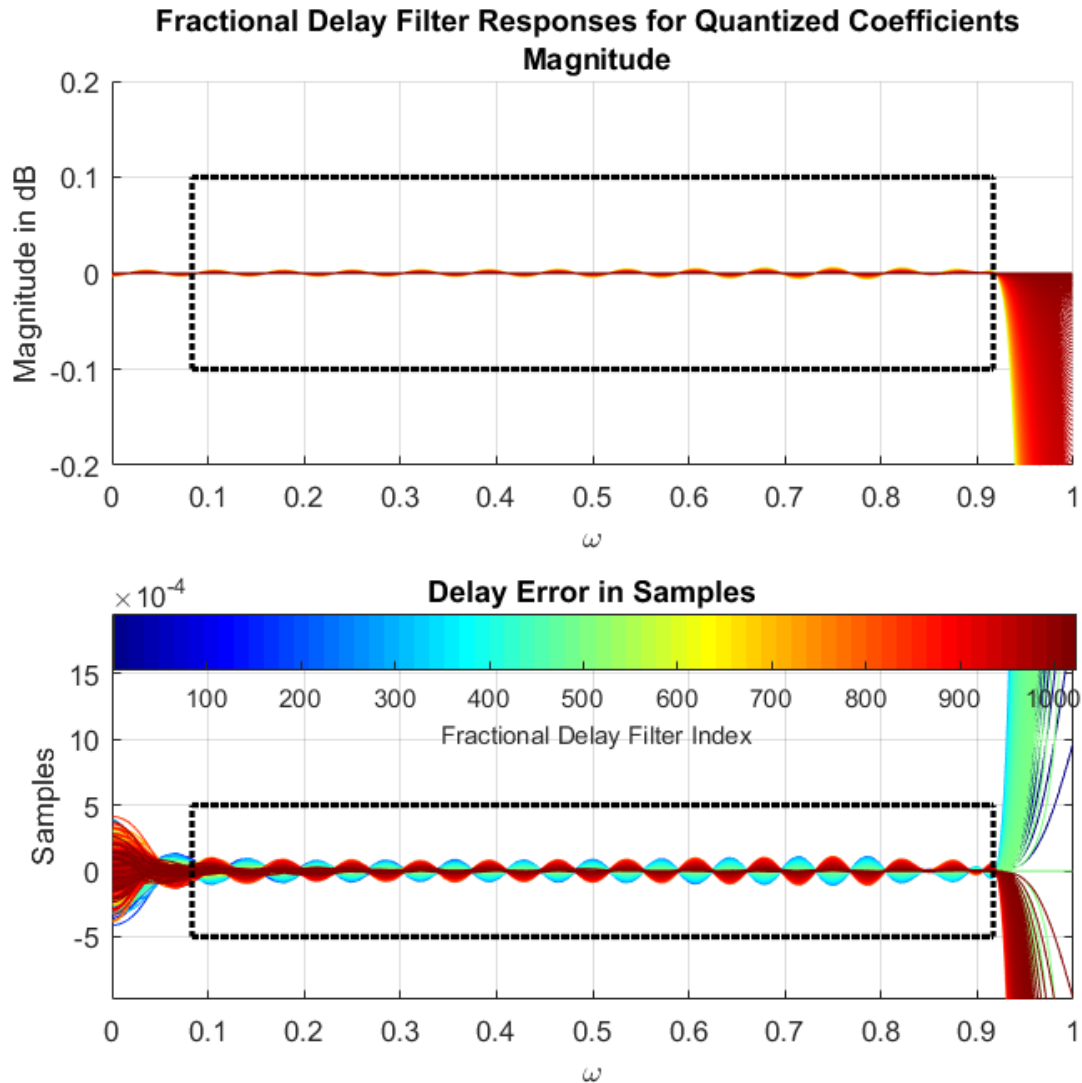

*Figure 7-3  Magnitude and delay error vs frequency for 56-tap filter with 19-bit tap coefficients. The required spectrally-pure region is indicated in bold dashed lines.*

---

[15] This is likely an enormous relief since post-correlation correction of this effect is problematic, not the least of which is due to RFI blanking.

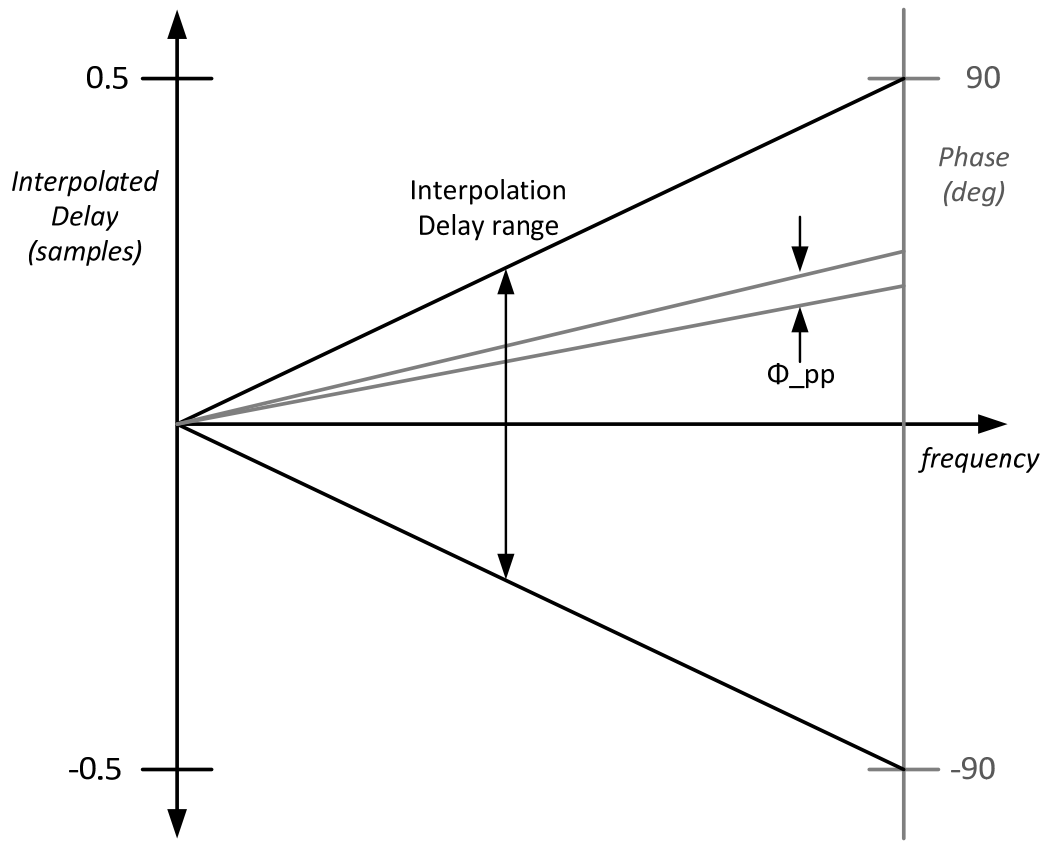

*Figure 7-4  Phase errors due to finite number of steps in delay interpolation between +/-0.5 samples. With 1024 steps, at the highest frequency the peak-to-peak phase deviation as delay sweeps between +/-0.5 samples, ϕ_pp, is π/1024.  This results in a worst-case coherence loss of ~0.0001% ($10^{-6}$) – nothing to worry about in sensitivity loss and highly unlikely needing any subsequent amplitude correction.*

The work in [1] left some uncertainty as to additional sensitivity losses due to re-sampling a coarsely-digitized signal.  Further modeling work reveals this is not of any significant concern, as the following plot, Figure 7-5 shows.  In this plot the blue trace is the sensitivity loss from cross-correlation of a wideband optimally quantized 4-bit signal.  The red curve is correlation of that signal, but first re-sampled to 8 bits and then correlated.  The sensitivity loss difference between them is ~0.0375%—due to the additional 8-bit post-resampling quantization.  Note that any frequency-dependent coherence loss due to a finite number of interpolation steps (Figure 7-4) is not apparent since it is several hundred times smaller than the offset between the red and blue curves.

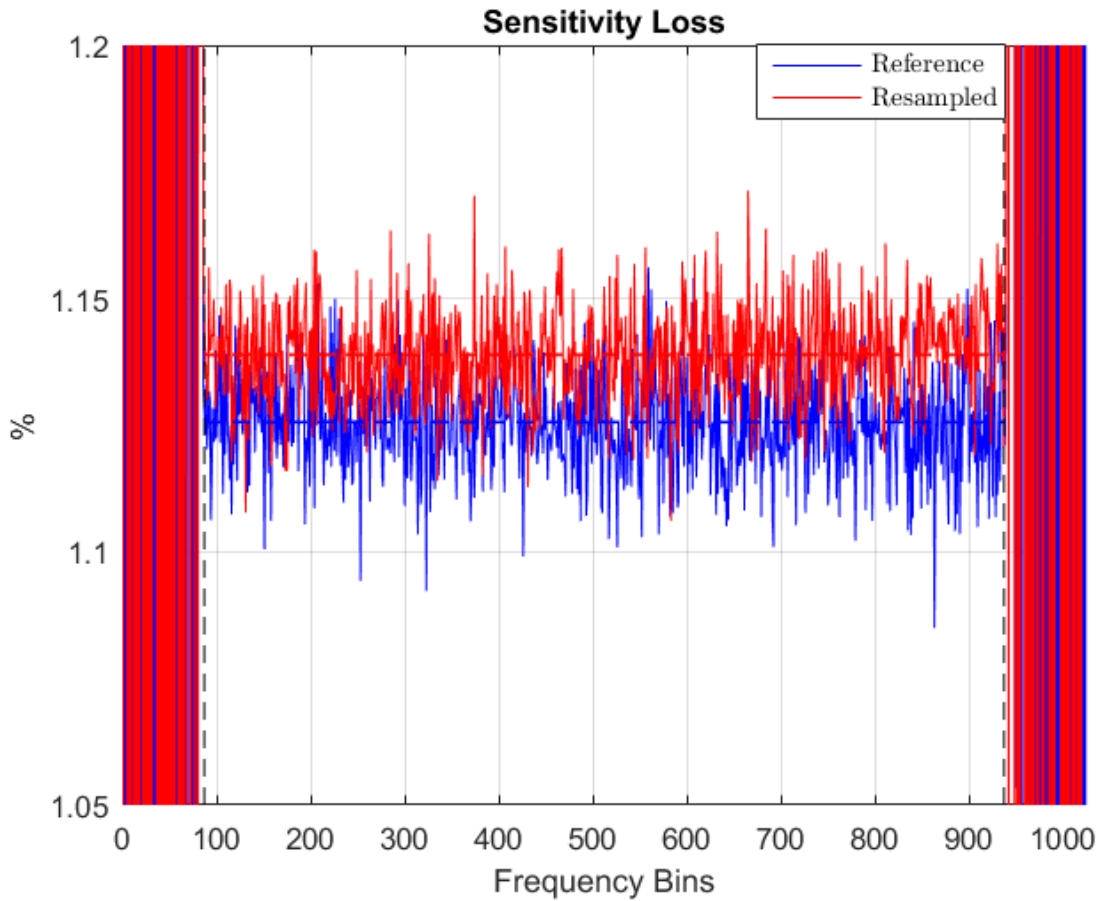

*Figure 7-5  Correlated sensitivity loss vs frequency for 4-bit sampling and correlation (blue) and 4-bit sampling, re-sampling, re-quantized to 8-bits, and correlation. This is for the case of 56 taps, 18-bit tap coefficients, and 1024 delay interpolation steps between +/-0.5 samples.  The additional coherence loss for re-sampling is the difference between the two traces, ~0.0375%.  Any frequency-dependent coherence loss due to the mechanism of Figure 7-4 is not apparent on this scale since it is several hundred times smaller than the separation between the blue and red traces.*

Finally, it is informative to see a time-domain trace of a 4-bit coarsely-quantized signal, the analog original, and the re-sampled 8-bit quantized signal.  This is shown in Figure 7-6, noting that the 8-bit re-sampled and re-quantized signal closely tracks the original 4-bit quantized signal and *not* the original analog signal since accomplishing the latter is fundamentally impossible as it would be creating information.

The 8-bit re-quantization trace does a sufficiently good job—with its finer level gradations—at faithfully re-producing the 4-bit originally-quantized signal and hence introduces a relatively small additional coherence loss, important for feasibility of the SCFO scheme.

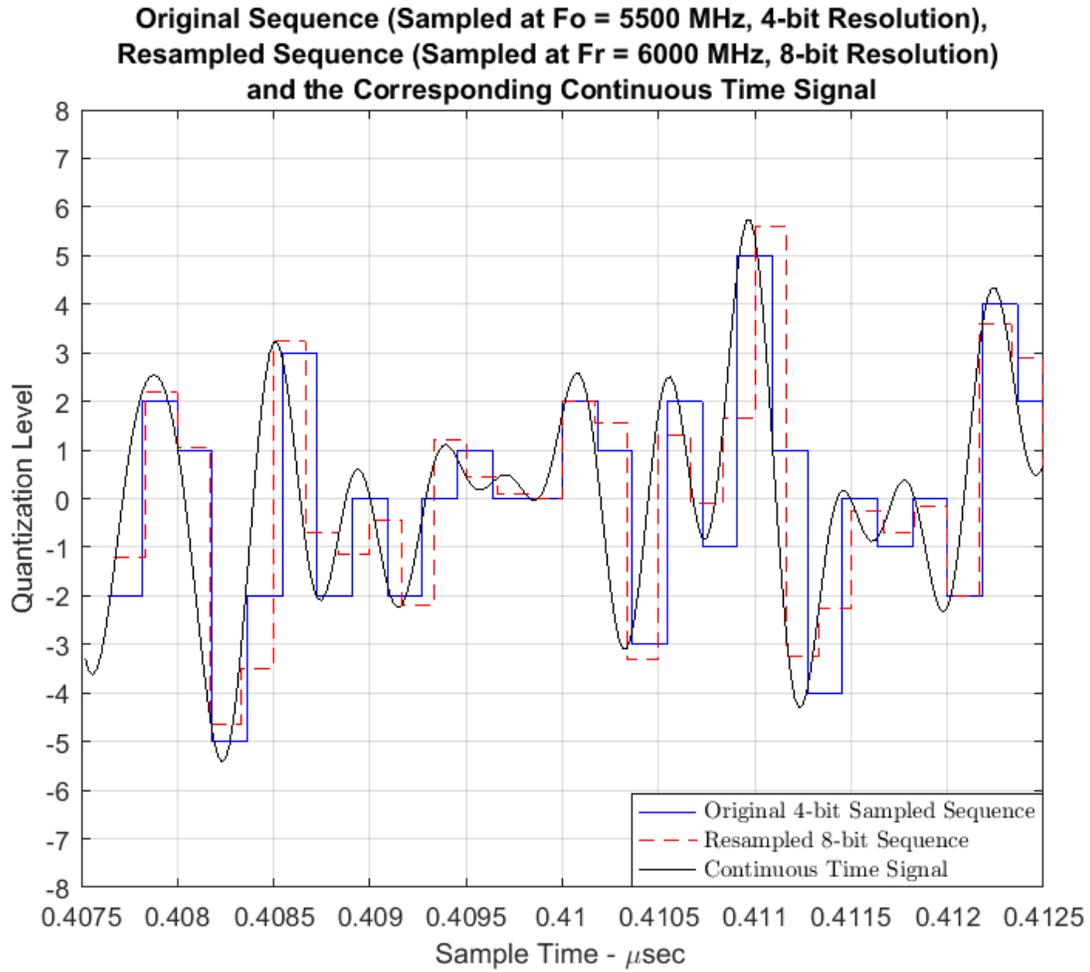

*Figure 7-6  Time domain display of traces showing original analog signal, 4-bit digitized representation, and 8-bit digitized re-sampled signal (i.e. derived from the 4-bit original digitized signal.) Sampling and re-sampling points are when state transitions occur.*

# 8 Scalability to SKA2

The primary issue regarding scalability to SKA2 is the sheer number of antennas (assume ~2000) and the need to maintain a minimum ~10 kHz baseline/differential frequency offset between any pair of antennas.

Under these conditions, the total range of frequency offsets spans ~2000 x 10 kHz = 20 MHz, or, +/-10 MHz.

While this does seem like a lot, even this is normally absorbed within the guard bands of each band (and is only of some concern when not Nyquist Zone 1 sampling).  Also, with the SCFO scheme, the guard bands are less rigidly demarcated than without it since aliasing into the guard band and even the pass band de-correlates, the main effect being additive noise (see section 3 Figure 3-7), which still puts some limits on aliasing so as to not degrade the SNR of the visibilities at the edges of the band.  This means that more of the guard band is spectrally pure, so loss of some overall total bandwidth due to SCFO is not an issue.

# 9   Possible Draw-backs and Issues of Concern

This section lists issues of concern and possible drawbacks.  These are believed to be as follows:

- The primary concern is that a high resolution synthesizer, be it DDS, fractional N, or other must be used in at least one location in SADT or DISH to generate the individual antenna sample frequencies and this synthesizer output contains jitter.  However, in the designs being considered, this is not believed to be an insurmountable problem.  Additionally, as SADT has carried out their design with the required assumption of sampler clock frequency offsets, there's no additional cost foreseen for SADT if the SCFO scheme is selected.

- The logic implementation of the re-sampler for Band 5 in CSP is reasonably straightforward as described in section 6.5 but it does consume power and costs money.  However, such a re-sampler is required for MeerKAT to SKA1 conversion and very likely VLBI re-sampling anyway and so is of no additional consequence.  As well, integrating all re-sampling, very fine delay tracking, and phase rotation into one block as illustrated in Figure 6-9 is architecturally attractive (simple) and results in superior delay and phase tracking with very likely no post-correlation correction requirements.

- The fact that the 1PPS presented to CSP essentially "jumps around" in time relative to the sampled data stream is something that must be carefully handled to ensure coherence for an observe block (a sequence of calibration and science observations forming a science project producing an image).  This is the case anyway without the SCFO scheme so is of no additional consequence.

# 10 Conclusions

This report has investigated the efficacy of the SCFO (Sample Clock Frequency Offset) scheme and finds that it is advantageous in rubbing out sample-clock dependent side-effects/self-interference in the correlator, and that such effects don't coherently sum in the beamformer. It has additional side-benefits of causing analog aliasing and aliased out-of-pass-band signals to de-correlate which can potentially be used to advantage in reduced DISH cost and power. Additionally, it is apparent that some cost and power savings in DISH may result since it means less stringent sample clock self-interference suppression measures (shielding, sample vs sub-sample clock distribution) need to be taken, particularly if the antenna-specific frequency $f_a$ is delivered by SADT to each antenna rather than the common frequency $f_c$.

SADT and DISH implementations were described and explored and there are believed to be no serious implementation issues of concern but further prototyping is required to verify that required clock frequencies and phase stability can be achieved in the critical (red) timing path indicated in Figure 2-1. It is accepted by SADT, DISH, and CSP that whether or not the SCFO scheme is employed, the 1PPS and the phase compensated clock at the antenna are not phase synchronous and this has telescope operational implications, namely that if an antenna power cycles during an observe block, a sky calibration is required before phase is re-established and visibilities on those baselines can be used.

CSP Mid.CBF implementation was explored in detail including fleshing out the re-sampler circuit, number of filter taps, re-sampler performance, subtle synchronization issues, and FPGA resource utilization and power estimation for the performance/power-driving Band 5 case. For the latter, ~20-50W (the exact number NDA protected) per antenna in Mid.CBF power is probably required for implementation, although such power is likely required anyway in VLBI mode for re-sampling to VLBI frequencies. If approved, Mid.CBF plans to use the re-sampling circuitry for very fine delay and earth-rotation phase corrections, which improves functionality and simplifies downstream signal processing. i.e. all such delay and phase effects are removed and need no further handling in downstream channelizers.

The bulk of re-sampling modelling is contained in [1], with additional results presented here indicating that re-sampling an initially 4-bit digitized signal, with 8-bit re-quantization after re-sampling, presents no significant additional coherence loss.

Finally, it would be prudent for an independent party to verify the effectiveness and implementation aspects of the SCFO scheme presented in this report in case some issues or effects were overlooked.

## 12  Acknowledgements

Thank-you to Dr. Thushara Gunaratne from NRC-Canada for doing all of the initial modeling investigation and work in [1] as well as further modeling work and plots presented in this report.  It would have been impossible to get this report complete without his high-energy effort.